\documentclass{aa}

\usepackage{natbib}
\bibpunct{(}{)}{;}{a}{}{,}

\usepackage{graphicx}

\usepackage{txfonts}
\usepackage[colorlinks=true,linkcolor=blue,citecolor=blue,filecolor=blue,urlcolor=blue]{hyperref}

\begin{document}

   \title{{{{Formamide formation on forsterite and Mg(OH)$_2$·H$_2$O}}}}

\author{        D. Campisi\inst{1,2,3,4}} \institute{ Current address: Aarhus Institute of Advanced Studies, Aarhus University, Høegh-Guldbergs Gade 6B, 8000 Aarhus C, Denmark \and Current address: Center for Interstellar Catalysis, Department of Physics and Astronomy, Aarhus University, 1520 Ny Munkegade 120, 8000 Aarhus C, Denmark \and Engineering Department, University of Perugia, Via Duranti 93, 06125 Perugia, Italy \and Institute for Theoretical Chemistry, University of Stuttgart, Pfaffenwaldring 55, Stuttgart, Germany  \\ \email{dario.campisi@aias.au.dk, campisi@theochem.uni-stuttgart.de, darcampisi@outlook.com} }

  \abstract 
   {Formamide formation in the Universe has been the subject of debate in several studies due to its importance as a precursor to amino acids and RNA. Current studies focus on the gas-phase and icy grain formation routes. However,  {formamide formation on bare silicate surfaces or hydrated
Mg-bearing molecular models remains largely unexplored in warm Mg-rich
dust environments, such as those potentially relevant to protoplanetary disks.}}
   {{{This study presents a {quantum-chemical investigation} of formamide formation on an Mg center and a silicate surface. {Mg(OH)$_2$·H$_2$O, considered here as a possible post-shock hydrated Mg-bearing species, and the bare (010) forsterite (Mg$_2$SiO$_4$) surface, {used as an analog of a bare dust-grain surface},} may promote formamide formation via NH$_2$ addition to H$_2$CO. {This may help clarify a potential additional pathway to formamide formation in warm Mg-rich environments.}}}}
   {I employed density functional theory (DFT), a quantum chemistry method, to study the addition of NH$_2$ to H$_2$CO {with an isolated Mg(OH)$_2$·H$_2$O molecular model} and on the forsterite (010) surface.
   First, I benchmarked various theoretical methodologies (M06-2X, PBE, PBE+U5, and VV) against the gold standard CCSD(T)-F12 and evaluated the performance of double-zeta and triple-zeta basis sets. I then examined the ability of Mg(OH)$_2$·H$_2$O and the forsterite (010) surface to {promote} formamide formation. {Finally, for the case with an entrance barrier, I computed unimolecular rate constants and half-lives to assess its kinetic behavior.}}
   {{Of the DFT methods I\ tested, M06-2X and PBE+U5 provide the closest agreement with CCSD(T)-F12 for the addition of NH$_2$ to H$_2$CO on an Mg center.} Mg(OH)$_2$·H$_2$O promotes proton transfer {within the} NH$_2$–CH$_2$O intermediate via the O–H group coordinated with Mg {following} NH$_2$ addition to H$_2$CO, {whereas on the forsterite (010) surface, a barrier-less addition pathway is found when NH$_2$ is initially physisorbed close to the formaldehyde center, followed by proton transfer leading to formamide formation.}}
   {{{This quantum-chemical study shows that the Mg-bearing model systems considered here can promote rapid proton transfer} following NH$_2$ addition to H$_2$CO, {whereas the corresponding bare gas-phase reaction proceeds through a large barrier for H abstraction to form formamide.}
   {For the Mg(OH)$_2$·H$_2$O model, the reaction step from the pre-reactive complex is predicted to occur on nanosecond timescales at temperatures approaching 300 K.} {The present work characterizes a potential pathway to formamide formation in Mg-bearing environments and provides molecular-level energetic and kinetic constraints for its future assessment in gas-grain astrochemical models.}}}


   \maketitle
\nolinenumbers

\section{Introduction}
\label{Introduction}

{The Universe serves as the cradle where organic molecules can form and develop, eventually becoming incorporated into planetary systems where life can begin \citep{Tielens_2013}. This is evident from the rich organic inventory found in carbonaceous chondrite meteorites; they are ancient materials formed under protoplanetary disk accretion and believed to have preserved the chemical conditions of the early Solar System \citep{Sephton_NPR_2002}. Isotopic analysis of these meteorites suggests that organic molecules likely have extraterrestrial origins \citep{Sandford_MPS_2001}.}
{Most of the organic precursors of life are thought to have originated under interstellar medium (ISM) conditions, where they are further chemically processed in protoplanetary disk environments, allowing them to be directly incorporated into forming planets \citep{Ehrenfreund:2000}.}

There are two hypothesized pathways for the formation of interstellar complex organic molecules, which are important precursors to building blocks of life such as amino acids. The "bottom-up" approach suggests that interstellar complex organic molecules form through the hydrogenation of CO on interstellar dust in the cold ISM, gradually building molecular complexity by adding a few atoms at a time. The "top-down" approach proposes that large carbonaceous molecules, such as polycyclic aromatic hydrocarbons (PAHs), fragment, releasing smaller molecules that can increase molecular complexity \citep{Tielens_2013}. {Finally, the products from the bottom-up and top-down processes are further processed during the evolutionary stages of the ISM, where they chemically evolve under protoplanetary disk and planetary conditions \citep{Ehrenfreund:2000}.}

{During the bottom-up phase,} the hydrogenation of CO leads to formaldehyde formation, which is known to be a reactant of the so-called Strecker amino acids reaction \citep{Aponte_2017}. However, for this reaction to occur, a high temperature and a liquid water environment, where equilibrium chemistry can occur, are necessary. So, it is unlikely that the Strecker reaction can occur in the ISM, where radical reactions are the driving forces of molecular complexity \citep{Rimola_Ugliengo_2010,Tielens_2013,Aponte_2017}.

Formamide (NH$_2$–HC=O) is a molecule that has been observed in star-forming regions \citep{Rubin_1971, Lopez_2019}, as well as in comets \citep{Bockelée-Morvan1997, Biver_2014, Goesmann_2015}, and is thought to act as a template for peptide bond formation and RNA synthesis \citep{Lopez_2019}.
Therefore, formamide plays a crucial role in understanding the origins of life.

Despite its detection, the mechanism of formamide formation is not yet fully understood, although it is believed to originate from the reaction of formaldehyde with the NH$_2$ radical \citep{Barone_2015}. Specifically, there is ongoing debate in the scientific community regarding whether formamide is formed in the gas phase \citep{Codella_2017, Vasyunin_2013, Barone_2015, Spezia_2016} or on dust surfaces \citep{Song_2016, Rimola_2018, Kan_2016}. \cite{Song_2016} studied the formation of formamide, which can occur efficiently through the hydrogenation of HNCO on water ice due to tunneling effects, while theoretical studies by \cite{Romero:2019} explored {the reaction} of NH$_2$ with HCO radicals on interstellar water ice. However, this reaction competes with the H transfer of HCO to NH$_2$, forming NH$_3$ instead.

 \cite{Kan_2016} experimentally studied the desorption of formamide as a product of cosmic ray interactions with ice. While this could explain the observed formamide in comets and circumstellar environments, the study did not address the mechanistic processes of its formation. Theoretical studies by \cite{Boland_2024} explored the formation of formamide from HCN and water ice surfaces through proton transfer effects. They suggest that a polar solid surface is necessary for abstracting the hydrogen atom, which requires overcoming large barriers. {The need to investigate the catalytic properties of a surface in promoting formamide formation is corroborated by the experimental studies by \cite{Douglas_2022}, which have shown that only a negligible amount of formamide is produced from the gas-phase reaction of NH$_2$ + CH$_2$O due to the presence of large barriers.} {Therefore, surface-mediated effects warrant further investigation. While most studies of formamide formation focus on ISM conditions, comparatively little work has explored reactions mediated by metal centers or bare silicate surfaces {in warm Mg-rich dust environments.}}

{Astrochemical models of protoplanetary disks provide an important large-scale context for the fate of formamide. In particular, \cite{Quenard_2018} show that formamide can be inherited from earlier evolutionary stages and released into the gas phase through thermal desorption in warm disk regions. This highlights inheritance and thermal desorption as possible contributors to gas-phase formamide in protoplanetary disks. At the same time, such models typically approximate dust grains as structurally uniform substrates governed by averaged binding energies and therefore do not resolve the elementary reaction steps occurring on specific mineral substrates. This leaves open the question of whether warm Mg-rich dust or hydrated Mg-bearing species can provide additional pathways for formamide formation or processing.}

{Such specific grain-surface processes may be relevant in protoplanetary disks, which} provide environments with moderate temperatures (tens to a few hundred kelvin) and relatively high densities (10$^8$–10$^{12}$ cm$^{-3}$ in the midplane), where grain-surface reactions {can} occur and where molecules formed or processed on dust grains {can} be incorporated into larger bodies such as asteroids and comets, ultimately contributing to planetary material.
{However, these environments are physically and chemically complex. The interplay of intense irradiation (particularly outside shielded regions), thermal desorption, accretion dynamics, and competing destruction processes ultimately determines whether a specific elementary reaction pathway is astrochemically relevant. Quantum-chemical data therefore represent valuable inputs to gas--grain astrochemical models for assessing the potential relevance of such pathways \citep{Henning:2013,Walsh:2014,Oberg:2023}.}

{Bare silicate surfaces and chondrite interplanetary dust particles (anhydrous surfaces) have been shown in past studies to exhibit catalytic activity in the Fischer-Tropsch reaction, which can occur in warm environments \citep{LLORCA:1998,Pareras:2024,Pareras:2023,Bradley:1994}}. {Alongside these known surface reactions}, observations suggest that formamide is also present in warm regions of the ISM, such as hot core regions, shocks, and protoplanetary disks \citep{Solomon:1973,Rubin_1971,Lopez_2019,Codella_2017,Adande:2013}. {However, the mechanism behind its formation remains a subject of debate, and {the possible role of Mg-assisted pathways on bare silicate surfaces or molecular Mg-bearing species in warm dust environments has not yet been characterized in detail.}}

{Forsterite (Mg$_2$SiO$_4$) is a silicate mineral belonging to the olivine family and is the main component of {dust grains}, in addition to being a major component of carbonaceous chondrite meteorites \citep{jones2007,Weinbruch:2000,Campisi_2022}.} The surface of forsterite has already been studied as a possible host of molecular species, such as PAHs, and a favorable site for catalytic reactions, such as the breaking of C-H bonds in PAHs due to the effect of polarization on the C-H bond \citep{Campisi_2022,Campisi_JPCA_2021,Campisi_MNRAS_2024}. {Hence, interesting chemistry might occur on forsterite surfaces.}

{On top of this, the action of shocks, originating from supernovae, stellar winds, or protoplanetary disk accretion, can sputter metal centers from dust grain surfaces and release them into the gas phase, where they could, in principle, catalyze important processes \citep{Shocks_1989,Shocks:2016}.} In this context, theoretical studies on metal ions—specifically Mg$^{2+}$—have shown a catalytic effect in reducing the barrier for the addition of N-bearing molecules to carbonyl compounds, lowering it by about 2 eV and thereby promoting the formation of peptide-like bonds \citep{Thripati:2021}.

{Mg(OH)$_2$·H$_2$O represents the smallest molecular unit of hydrated brucite that could form through the interaction of water with magnesium-bearing surfaces during shock events in protoplanetary disks and planetary system conditions as supported by simulations \citep{Cazaux:2019} and by detections in carbonaceous chondrite objects \citep{MACKINNON1979, Milliken2009, MacLennan2024}.}
Broad D/H variations in LL3 pyroxene chondrules further suggest {a hydration process involving} precursor phyllosilicates in circumstellar or interstellar environments \citep{Hofmeister:2006,DELOULE19983367}. 
{However, no studies have yet investigated how Mg(OH)$_2 \cdot$ H$_2$O promotes the NH$_2$ + H$_2$CO reaction pathway.}

{In this study, I present quantum-chemical data for the formation of formamide via the reaction of NH$_2$ with H$_2$CO on the forsterite (010) surface. Additional calculations on an isolated Mg(OH)$_2 \cdot$ H$_2$O molecular model are included to provide insight into alternative reaction pathways. Importantly, Mg(OH)$_2 \cdot$ H$_2$O is proposed here as a possible post-shock hydrated Mg-bearing species, motivating future observational searches aimed at its identification and characterization. Its inclusion is therefore solely exploratory, and I\ aim to probe how such a hydrated Mg center promotes the NH$_2$ + H$_2$CO reaction pathway. By characterizing the elementary energetics, transition states (TSs), and unimolecular rate constants of these Mg-bearing model pathways, this work provides molecular-level constraints for their future assessment in astrochemical models under specific local conditions.}

{Specifically, I examined the theoretical methods that might accurately describe the formation of formamide on a Mg(OH)$_2$·H$_2$O molecule (Sect. \ref{benchmark}). Upon choosing the methodology (described in Sect. \ref{Simulations}) and characterizing the potential energy surface (PES) of the addition of NH$_2$ onto the formaldehyde–Mg(OH)$_2$·H$_2$O complex, I studied how the same process compares when it occurs on a periodic (010) forsterite surface (Sect. \ref{Formamide_formation}).} {Finally, I discuss the potential astrophysical implications of these pathways (Sect. \ref{Astrophysical Implications}).

\section{Simulations}
\label{Simulations}

{To study the reaction mechanism of NH$_2$ with formaldehyde in Mg-bearing model systems, I modeled formaldehyde adsorbed on Mg(OH)$_2$·H$_2$O and on the (010) surface of forsterite (Fig. \ref{fgr:cluster}), {which has been reported as the most stable forsterite surface} \citep{deLeeuw:2001, Bruno2014, Zamirri:2017}}. Specifically, I used the ORCA code \citep{ORCA,ORCA_update} to investigate the reaction on Mg(OH)$_2$·H$_2$O and the SIESTA code \citep{SIESTA} for calculations on both the Mg(OH)$_2$·H$_2$O model and the periodic (010) forsterite surface.}

{For the Mg(OH)$_2$·H$_2$O model, I benchmarked the {Minnesota} M06-2X \citep{Zhao2008}, {Perdew--Burke--Ernzerhof} PBE  \citep{PBE}, PBE+U5 \citep{LDA+U}, and {Vydrov--Van Voorhis} VV \citep{VV} exchange-correlation functionals against {the explicitly correlated coupled-cluster
singles and doubles with perturbative triples} CCSD(T)-F12 \citep{CCSDT-F12}.} Double-zeta and triple-zeta basis sets with polarization functions, pcseg-1 and pcseg-2 respectively \citep{Jensen2014}, were utilized for M06-2X calculations using the ORCA code, along with auto-generated auxiliary basis sets \citep{Stoychev2017}. The resolution of identity approach was applied for the Coulomb integral in the density functional theory (DFT) calculations \citep{RI} to reduce the computational cost.
For the CCSD(T)-F12 method, I employed the {F12-optimized correlation-consistent polarized valence double-zeta cc-pVDZ-F12} basis set \citep{Kirk:2008} with the {complementary auxiliary basis set  cc-pVDZ-F12-CABS}  and {correlation-fitting cc-pVTZ/C} auxiliary basis set for resolution of identity calculations, also using the ORCA code.

For the PBE, PBE+U5, and VV functionals, numerical atomic orbitals were employed with  {double-zeta polarized (DZP) and triple-zeta polarized (TZP)} basis sets and Troullier–Martins norm-conserving pseudo-potentials \citep{Pseudo:1991}. A split-norm radius of 0.50 {was} used for hydrogen atoms, while a split-norm radius of 0.30 {was} applied for other atoms in the TZP basis sets, using the SIESTA code.

The choice of functional sets here is carefully considered to balance accuracy and computational cost. Specifically, M06-2X is a {hybrid} {meta-generalized gradient approximation (meta-GGA)} functional that has been widely employed and benchmarked in previous studies \citep{Mardirossian2016,Jensen:2019}, ensuring a good description of neutral-neutral reactions involving radicals in a doublet spin state. PBE, a GGA functional, is chosen as it is the standard for periodic calculations \citep{PeriodicDFT,Campisi_JPCA_2021}, which cannot feasibly employ hybrid functionals or meta-GGAs due to their computational expense. In the past, PBE alone has been shown to accurately describe the forsterite surface properties and structure as well as adsorption and proton-transfer processes involving radicals and organic molecules on the forsterite surface, yielding results in good agreement with M06-2X calculations \citep{Andersen:2023,Campisi_JPCA_2021,Campisi_2022,Campisi_MNRAS_2024,Demichelis:2015}. Here, to enable a fair comparison between the level of theory used for the reaction occurring in the isolated Mg(OH)$_2$·H$_2$O model and that on the forsterite surface, I included PBE+U5, a Hubbard-corrected PBE functional {in which the $U$ correction is applied only to the nitrogen $2p$ orbitals of the NH$_2$ radical to account for strong on-site Coulomb interactions} \citep{Himmetoglu2014}. This approach was adopted to study the formation of the C--N bond, where correlation effects may be underestimated when using PBE alone. {In this work, I employed $U=5$ eV, which was} selected based on its agreement with the CCSD(T)-F12 calculations, as discussed in Appendix D. {This choice allows the} PBE+U5 description of the reaction on the periodic surface to be compared with the M06-2X description of the reaction in the isolated Mg(OH)$_2$·H$_2$O model. {This choice is also consistent with previous applications of Hubbard corrections to localized $p$ states. In particular, $U$ values in the range of 5--7 eV have been employed for N $2p$ states in nitrogen-bearing systems \citep{Pardo:2008,Morbec:2016,Wanxue:2018,Denchfield:2025}. Pronounced electronic localization is not exclusive to $d$ and $f$ states but has also been investigated in $p$ states in other systems, for which Hubbard corrections have been applied \citep{Campisi_JPCA_2021,Campisi_2022,Macke:2024}.} In addition, the VV functional, a vdw-DF2 functional with an optB88 exchange-correlation function that includes self-consistent nonlocal dispersion corrections \citep{Klime_2010}, is chosen to evaluate the effect of dispersion in the reaction pathway.

In ORCA, the threshold for {the self-consistent field  (SCF)} convergence {was} set very tight, with values below 1.0 × 10$^{-9}$ a.u., while geometry optimization {was} stopped when the maximum gradient {was} below 0.0003 hartree/bohr. Harmonic vibrational frequencies {were} computed to ensure proper characterization of saddle points (with only one imaginary frequency) and minima (no imaginary frequency). 

Minima {were} optimized using the limited-memory Broyden–Fletcher–Goldfarb–Shanno algorithm \citep{Liu1989}, and the TS structures {were} optimized using the eigenvalue-following TS finding algorithm implemented in ORCA. {Using the SIESTA code, relaxed scan optimizations were performed with the atomistic simulation environment (\citealt{Larsen_2017}) by varying the relevant bond length. When the scan indicated an energy barrier, the resulting structure was used as the initial guess for TS optimization with the Sella algorithm \citep{Sella}.} When the minimum energy {pathway} could be described by a single bond length {coordinate}, this approach was sufficient. In other cases, or when the scan suggested a barrier-less process and further verification was required, the relaxed {scans} were used to initialize a climbing-image nudged elastic band (CI-NEB) {calculation} \citep{NEB:2000} {to verify whether a TS was present}. {Any TS candidate identified by CI-NEB was subsequently optimized with Sella. For the Sella optimizations, a partial Hessian including the atoms directly involved in the bond-forming process was employed. Convergence was defined as a maximum force below 0.01 eV/$\AA$ for the relaxed scans and Sella optimizations and below 0.05 eV/$\AA$ for CI-NEB.}

\begin{figure}
  \caption{Periodic model of the forsterite (010) surface employed in this work at different orientations (A, B, and C) represented by the Cartesian axes. The surface vacuum region is along the z-axis.}
\includegraphics[width=\hsize]{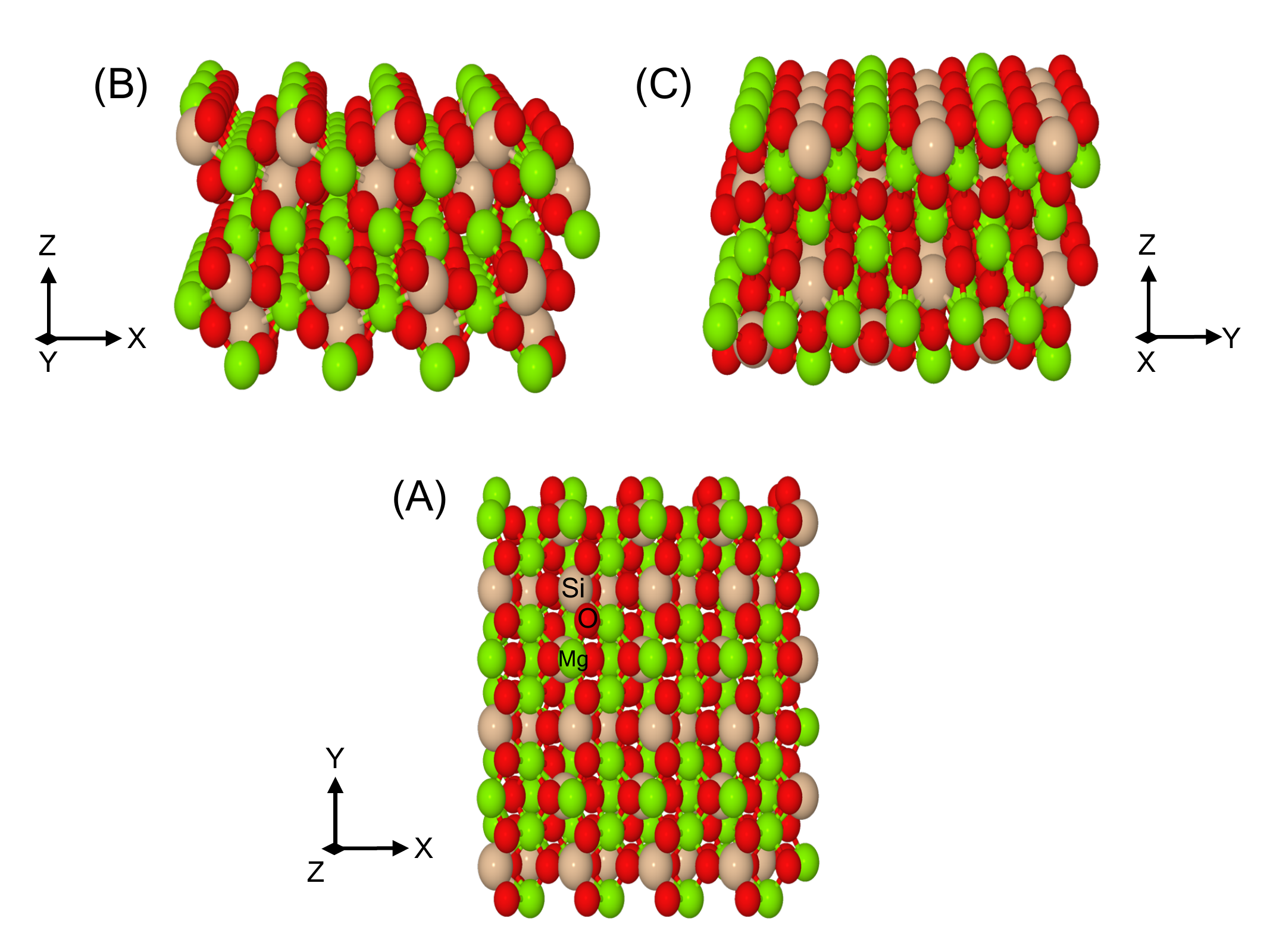}
  \label{fgr:forsterite}
\end{figure}

\begin{figure}
  \caption{{Schematic top views of the periodic forsterite (010) surface and the isolated molecular Mg(OH)$_2$·H$_2$O model, with atomic labels. For the periodic surface model, the vacuum region extends along the z-axis, as indicated by the Cartesian axes.}}
\includegraphics[width=\hsize]{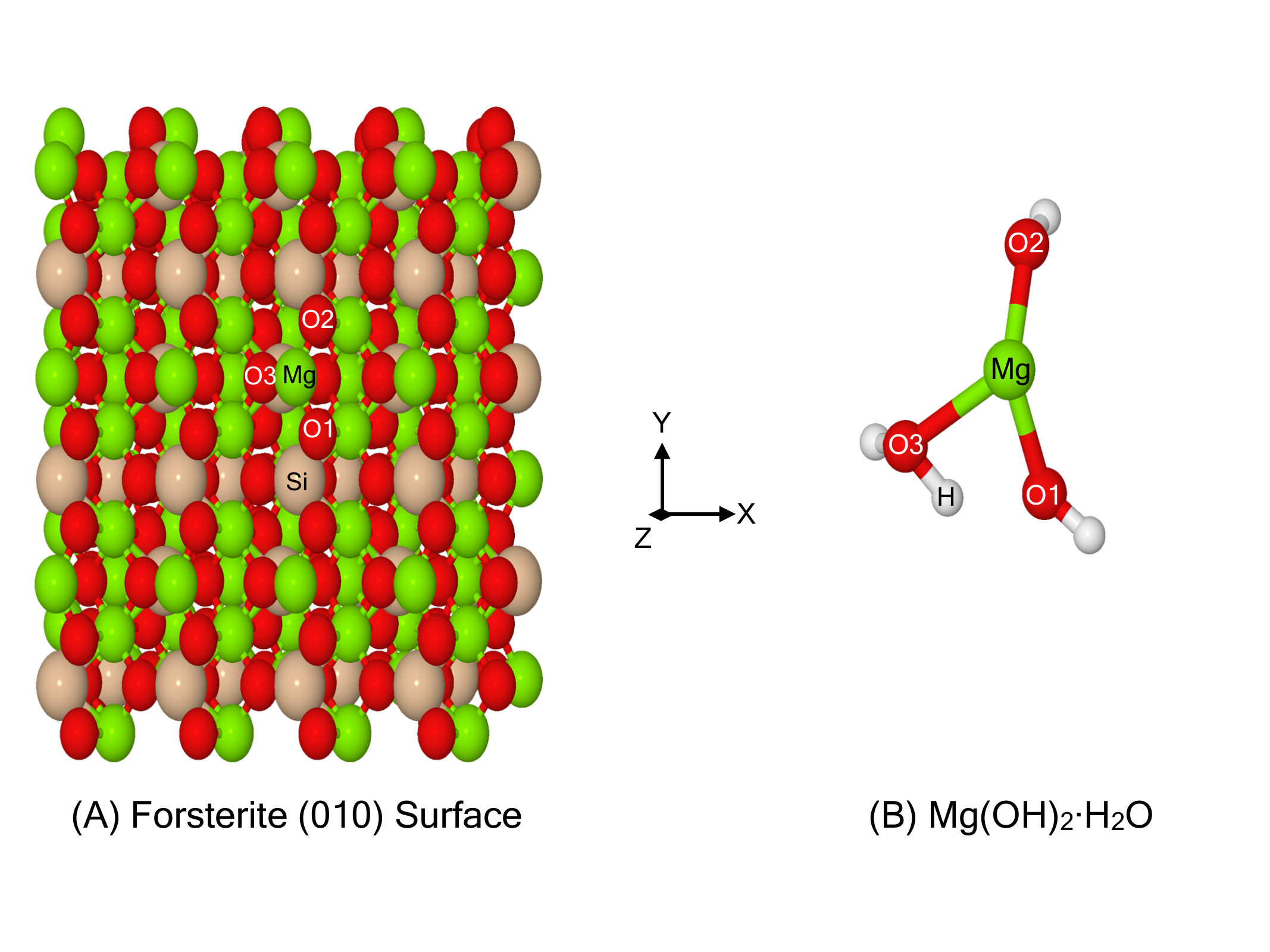}
  \label{fgr:cluster}
\end{figure}

Using the SIESTA code, the Mg(OH)$_2$·H$_2$O model {was} constructed in a 30 × 30 × 30~\AA~cubic unit cell {to mimic isolated-molecule conditions}. This setup allowed for comparable results with calculations performed using the ORCA code. The periodic (010) forsterite surface, {shown in Fig.~\ref{fgr:forsterite}}, {was} modeled in a 19.00 × 17.94 × 35.20~\AA~super-cell, as optimized in a prior study \citep{Campisi_JPCA_2021}, which included a vacuum region of 25~\AA~to prevent interactions between periodic images. The dimensions of the unit cell for both Mg(OH)$_2$·H$_2$O and the (010) forsterite surfaces are chosen to ensure that reactants (e.g., formaldehyde and NH$_2$) would not interact with their periodic images. {Given the large unit cells, all calculations employed a 1 x 1 x 1 Monkhorst-Pack k-points grid \citep{Kpoint} to sample the Brillouin zone.}

The real-space grid uses a plane-wave energy mesh cut-off of 200 Ry. This mesh cut-off is not related to the basis sets and does not affect the accuracy of the calculation; it is set to limit the egg-box effect and ensure smoother optimization convergence \citep{SIESTA}. During the optimization procedure, an electric field {was} applied to correct for the resultant dipole moment along the z-axis.

The optimization is considered converged when the forces are below 0.005~eV$/$\AA~for the Mg(OH)$_2$·H$_2$O model and 0.01~eV$/$\AA~for the periodic model. Geometry optimization {was} carried out using the limited-memory Broyden–Fletcher–Goldfarb–Shanno algorithm implemented in the atomistic simulation environment with SIESTA attached as a calculator \citep{Larsen_2017}. For the periodic model, I computed partial harmonic vibrational frequencies, focusing only on the atoms involved in the reaction (formaldehyde and NH$_2$) to reduce the computational cost of calculating the Hessian for the periodic structure.

{The energies of the optimized stationary points computed with ORCA and SIESTA were corrected a posteriori for {the basis set superposition error (BSSE)} using the counterpoise method \citep{Boys01101970,Sordo1988}. Because this correction is defined with respect to separated reactant fragments, applying it independently to the CI-NEB images could introduce artificial variations along the reaction profile. Therefore, the CI-NEB profiles were referenced to the preceding intramolecular intermediate to avoid introducing geometry-dependent a posteriori BSSE corrections along the reaction path. All calculations employed the unrestricted spin formalism.}

Binding energies are defined as \begin{eqnarray} E_\text{b} & = & E_\text{r} - E_\text{p} \label{eq1} ,\end{eqnarray} where $E_\text{b}$ is the binding energy, $E_\text{r}$ is the energy of the reactants, and $E_\text{p}$ is the energy of the product or intermediate. Therefore, positive values of $E_\text{b}$ indicate a favorable process. Reaction energies are, by definition, the opposite (negative sign) of binding energies. Non-covalent interaction is computed as 

\begin{eqnarray} E_\text{vdw} & = & E_\text{v} - E_\text{r} \label{eq2} ,\end{eqnarray} 

{\noindent where $E_\text{vdw}$ is the energy of a non-covalent interaction in a van der Waals (vdW) complex, $E_\text{v}$ is the energy of the vdW complex, and $E_\text{r}$ is the energy of the reactants. Positive values of $E_\text{vdw}$ indicate an unfavorable non-covalent interaction. Energy barriers are computed as 
\begin{eqnarray} E_\text{a} & = & E_\text{TS} - E_\text{i} \label{eq3} ,\end{eqnarray} 
where $E_\text{a}$ is the energy barrier, computed as the difference between the energy of the TS ($E_\text{TS}$) and the energy of the intermediate ($E_\text{i}$) that precedes the TS.}

\section{Results}
\label{Results}

{To assess how Mg-bearing environments promote the reaction of NH$_2$ with formaldehyde (H$_2$CO), I first benchmarked the selected DFT approaches against the CCSD(T)-F12 reference for the isolated Mg(OH)$_2$·H$_2$O model (Sect. \ref{benchmark}). I then compared the resulting reaction pathway with the pathway obtained on the periodic forsterite (010) surface (Sect. \ref{Formamide_formation}).}

\subsection{{Mg(OH)$_2$·H$_2$O-assisted formamide formation}}
\label{benchmark}

\begin{figure*}
  \caption{{Schematic representation of the PES for the reaction of NH$_2$ with H$_2$CO leading to formamide in the presence of the isolated Mg(OH)$_2$·H$_2$O model. Relative energies obtained at the indicated levels of theory are reported in the same order as the corresponding labels in the legend and are referenced to the separated reactants. All stationary-point geometries were optimized at the M06-2X/pcseg-1 level of theory. The energies reported for the other electronic-structure methods were obtained from single-point calculations on the corresponding M06-2X/pcseg-1 optimized geometries. For the first and second TS, the reported values are barrier heights relative to the preceding minimum.}}
  \includegraphics[width=\hsize]{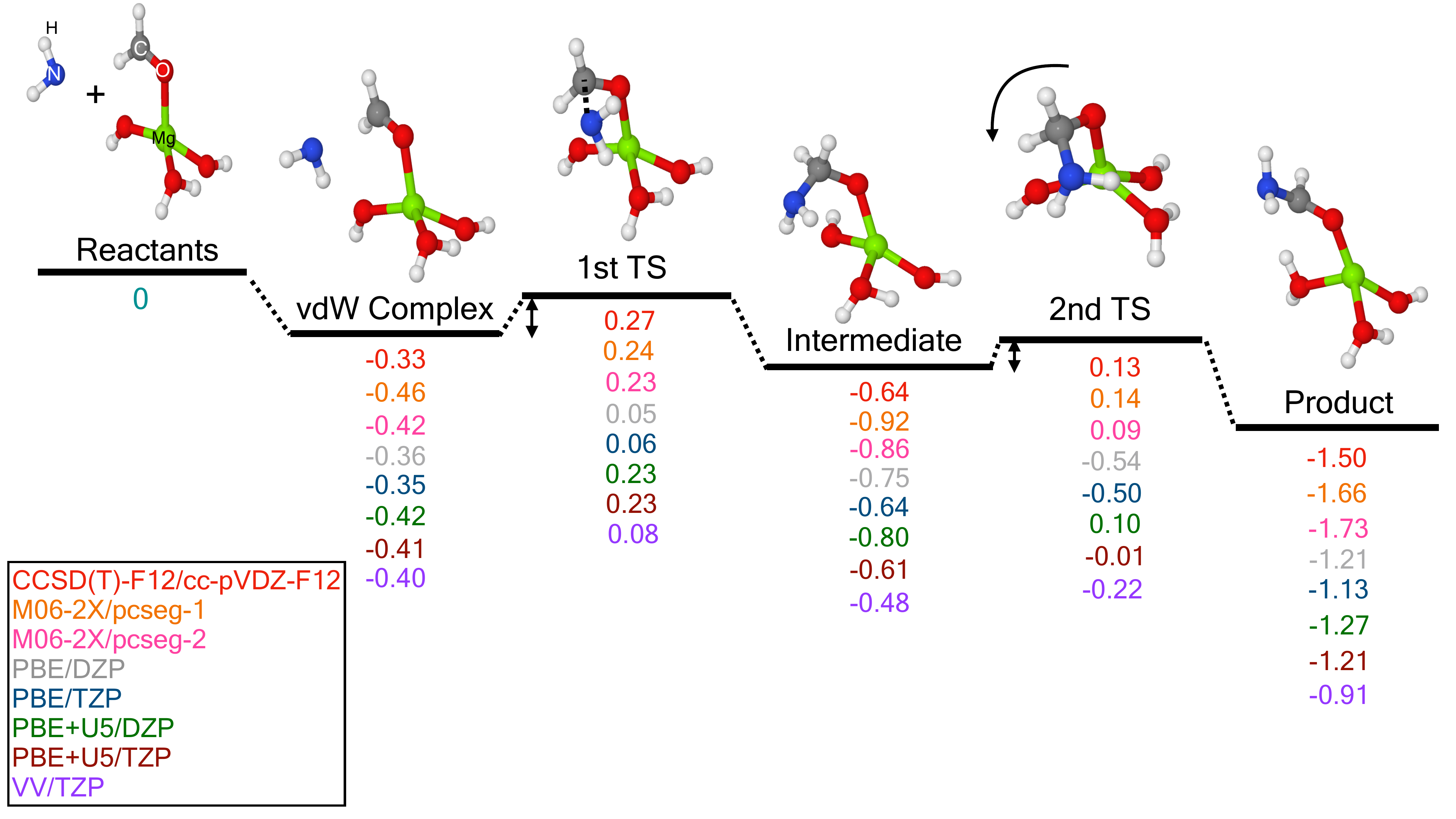}
  \label{fgr:PES}
\end{figure*}

{Figure \ref{fgr:PES} shows the PES for the formation of formamide coordinated to the Mg(OH)$_2$·H$_2$O molecular model, calculated using various levels of theory.} When NH$_2$ approaches formaldehyde, before crossing the first energy barrier, it forms a non-covalent (vdW) complex with an H-bond between the N atom and the O atom of the Mg(OH)$_2$·H$_2$O model.

{The gold standard CCSD(T)-F12 method predicts a favorable and weak formation energy of -0.33 eV, which serves as the reference point for comparison. The M06-2X/pcseg-1 method shows a slightly stronger interaction with a formation energy of -0.46 eV, with a relative error of 0.39 with respect to CCSD(T)-F12, while M06-2X/pcseg-2 predicts -0.42 eV, with a relative error of 0.27 with respect to CCSD(T)-F12. Both of these methods predict stronger interactions than CCSD(T)-F12, with M06-2X overestimating the interaction energy. The PBE/DZP method shows a formation energy of -0.36 eV, with a relative error of 0.09 from CCSD(T)-F12, suggesting a reasonable agreement with the reference value. The PBE/TZP method gives a value of -0.35 eV, with a relative error of 0.06, indicating even better agreement with the CCSD(T)-F12 result. The PBE+U5/DZP and PBE+U5/TZP methods estimate similar energy values of -0.42 eV and -0.41 eV, respectively, with relative errors of 0.27 and 0.24 from CCSD(T)-F12, suggesting that the inclusion of the Hubbard U correction does not significantly affect the overall non-covalent formation energy. Finally, the VV/TZP method yields a non-covalent formation energy of -0.40 eV, with a relative error of 0.21 from CCSD(T), indicating a reasonably close agreement, though still slightly larger than CCSD(T)-F12.}

{Once the vdW complex is formed, only M06-2X and PBE+U5 predict barriers that are comparable to CCSD(T)-F12, with close agreement corresponding to relative errors of about 0.11–0.15. In contrast, PBE and VV underestimate the barrier, with relative errors of about 0.70–0.81.}

{After overcoming the first barrier, the intermediate NH$_2$CH$_2$O is formed by the addition of NH$_2$ to the carbon atom of formaldehyde. Based on CCSD(T)-F12 calculations, the formation of this intermediate is exoergic and favorable. M06-2X shows an overestimation, compared to CCSD(T)-F12, of the intermediate formation energy with relative errors of 0.44 for the pcseg-1 basis set and 0.34 for the pcseg-2 basis set. PBE exhibits a relative error of 0.17 for the DZP basis set, while the TZP basis set aligns perfectly with CCSD(T)-F12. PBE+U5 shows a relative error of 0.25 with the DZP basis set and a negligible relative error of 0.05. VV demonstrates a larger relative error of 0.25 compared to CCSD(T)-F12.}

{After formation of the intermediate, rotation around the O-C bond reorients the intermediate toward the OH group of Mg(OH)$_2$·H$_2$O. This conformational rearrangement corresponds to the second TS. After the rotational barrier is overcome, proton transfer from the carbon atom of the intermediate to the oxygen atom of the O-H group proceeds without an additional resolved barrier, as shown by the CI-NEB calculation (see Fig. \ref{fgr:ProtonTransfer}).}

{The second rotational barrier has a value of 0.13 eV based on CCSD(T)-F12 calculations. M06-2X/pcseg-1 and PBE+U5/DZP predict barriers of 0.14 and 0.10 eV, respectively, corresponding to relative errors of 0.08 and 0.23 with respect to CCSD(T)-F12. Owing to the small magnitude of the reference barrier, the relative error of 0.23 for PBE+U5/DZP corresponds to an absolute deviation of only 0.03 eV. In contrast, M06-2X/pcseg-2 predicts a lower barrier of 0.09 eV, with a relative error of 0.31, while PBE+U5/TZP does not reproduce the barrier and yields a value of -0.01 eV. The lowering of this shallow rotational barrier observed when moving from pcseg-1 to pcseg-2 and from DZP to TZP indicates that this feature is particularly sensitive to the basis set. PBE without the Hubbard correction and VV/TZP also fail to reproduce the rotational barrier, yielding negative values. These results show that, at the DZP level, the Hubbard correction improves the PBE description not only of the first C--N bond-forming barrier, but also of the subsequent rotational step.}

{After overcoming the rotational barrier, the formation of formamide is highly favorable and exothermic, with a reaction energy of approximately -1.50 eV according to CCSD(T)-F12 calculations. The M06-2X method overestimates the product formation energy with a relative error of 0.10 for the pcseg-1 basis set. In contrast, M06-2X/pcseg-2 shows a slightly larger relative error of 0.18. The PBE functional underestimates the formation energy with relative errors of approximately 0.19 with the DZP basis set and 0.25 with the TZP basis set. Similarly, PBE+U5 underestimates the formation energy with relative errors of about 0.15 with DZP and 0.19 with TZP. Overall, PBE+U5 and M06-2X exhibit comparable magnitudes of relative error, with the former underestimating the formation energy and the latter overestimating it. Finally, the VV method has the largest relative error, 0.39.}

{Overall, M06-2X/pcseg-1 and PBE+U5/DZP provide the closest agreement with the CCSD(T)-F12 potential energy profile among the tested DFT approaches. In particular, relative to PBE/DZP, the Hubbard correction improves the description of the overall reaction profile, including both the first C--N bond-forming barrier and the second rotational barrier. These results support the use of PBE+U5/DZP to compare the isolated Mg(OH)$_2$·H$_2$O model with the periodic forsterite (010) surface.}

\subsection{{Mg(OH)$_2$·H$_2$O versus forsterite (010)}}
\label{Formamide_formation}

{Table \ref{tab:Formamide_formation} reports the counterpoise- and zero-point-corrected relative energies and barrier heights for the reaction of NH$_2$ with formaldehyde in the isolated Mg(OH)$_2$·H$_2$O molecular model and along the most favorable pathway on the periodic forsterite (010) surface. On forsterite, this pathway corresponds to a Langmuir--Hinshelwood-type local reaction in which H$_2$CO is chemisorbed and NH$_2$ is initially physisorbed in close proximity to it.}

{The PES for the addition of
NH$_2$ to formaldehyde adsorbed on the periodic forsterite (010)
surface is shown in Fig.~\ref{fgr:periodic_PES}. Three
distinct surface-reaction pathways were examined. In the first
pathway, gas-phase NH$_2$ directly attacks chemisorbed H$_2$CO. This
corresponds to an Eley--Rideal-like mechanism and presents an
activation barrier of 0.79 eV. In the second pathway, H$_2$CO is
chemisorbed on the surface, while NH$_2$ is initially physisorbed in
close proximity to it. Since both reactants are adsorbed before the
reaction, this configuration represents a local
Langmuir--Hinshelwood-type reaction limit. From this pre-reactive
configuration, the relaxed scan along the C--N distance does not reveal
an intervening maximum between the pre-reactive configuration and the
formation of the chemisorbed C--N-bonded intermediate
(Fig.~\ref{fgr:N-C_Scan}). Barrier-less evolution from a
physisorbed precursor to a chemisorbed state is a recognized limiting
case of physisorption-to-chemisorption potential-energy profiles in
surface chemistry \citep{Huber:2019,Wang:JPCC:2007}.} 

{In the third pathway, both reactants are surface-bound, with NH$_2$ chemisorbed on a neighboring Mg atom. This second Langmuir--Hinshelwood-type pathway requires overcoming a rotational barrier of 0.94 eV, followed by a C--N addition barrier of 0.90 eV. A complete kinetic treatment of the Langmuir--Hinshelwood mechanism, including long-range surface diffusion, surface coverage, and encounter probabilities, is beyond the scope of the present calculations.}

{In all three pathways, once the C--N bond is formed, the surface-mediated H-transfer step leading to chemisorbed formamide proceeds without an additional resolved barrier. The most favorable pathway is therefore the Langmuir--Hinshelwood-type configuration in which NH$_2$ is weakly physisorbed close to chemisorbed H$_2$CO. This pathway exhibits a reaction sequence similar to that obtained with the isolated Mg(OH)$_2 \cdot$ H$_2$O model, with the important difference that C--N bond formation on the surface is barrier-less.} {The barrier-less C--N bond formation in this pathway appears to be favored by the orientation of the weakly physisorbed NH$_2$ radical relative to chemisorbed H$_2$CO on the surface, as discussed in detail in Appendix B.}

NH$_2$ addition to {formaldehyde} in a pure gas phase medium {(in the absence of metal centers)}, as reported in the literature \citep{Rimola_2023}, shows an energy barrier of 0.23 eV relative to the vdW complex at the level of CCSD(T)//M06-2X-D3. This is comparable to the value I found for Mg(OH)$_2$·H$_2$O at the CCSD(T)-F12/M06-2X level, with a difference of only 0.03 eV. A significant difference arises when computing the barrier from the reactants at infinite distances (the asymptote). In this case, the barrier is 0.08 eV at the CCSD(T)-F12/M06-2X level {for} Mg(OH)$_2$·H$_2$O and 0.19 eV at the CCSD(T)//M06-2X-D3 {level} in the gas phase. {The lower barrier relative to the separated reactants in the Mg(OH)$_2\cdot$H$_2$O pathway arises because its pre-reactive vdW complex is more strongly stabilized than the corresponding gas-phase NH$_2\cdots$H$_2$CO complex, which lies only 0.04~eV below the separated NH$_2$ and H$_2$CO reactants at the CCSD(T)//M06-2X-D3 level.}

{This comparison shows that Mg(OH)$_2$·H$_2$O does not significantly lower the C--N bond-forming entrance barrier relative to the reaction in the absence of a metal center. Its main effect is instead observed after NH$_2$ addition, where rotation of the intermediate toward the Mg-coordinated OH group is followed by H transfer without an additional resolved barrier, leading to the formation of a second water molecule coordinated to Mg (Fig. \ref{fgr:PES} and Fig. \ref{fgr:ProtonTransfer}). By contrast, for the corresponding reaction in the absence of a metal center, H loss from the C--N-bonded intermediate to form formamide has a reported intrinsic barrier of 0.47~eV \citep{Rimola_2023}.}

{On the periodic forsterite (010) surface, the most favorable Langmuir--Hinshelwood-type configuration promotes a barrier-less pathway for C--N bond formation, and the subsequent surface-mediated H-transfer step also proceeds without a resolved barrier. The relative energies of the final product differ from the corresponding Mg(OH)$_2$·H$_2$O values by 0.09 eV, indicating comparable stabilities.}

\begin{table*}
    \centering
\caption{Counterpoise- and zero-point-energy-corrected relative energies and barrier heights for NH$_2$ addition to H$_2$CO on Mg(OH)$_2\cdot$H$_2$O and the periodic forsterite (010) surface.}
    \begin{tabular}{@{}cccccc@{}}
        \hline\hline
         {Model} & {vdW} & TS1 & Int. & TS2 & Prod.  \\
        \hline
        {Mg(OH)$_2\cdot$H$_2$O [M06-2X/pcseg-1]} & -0.35 & 0.25 & -0.63 & 0.03 & -1.43 \\
        {Forsterite (010) [PBE+U5/DZP]} & {-0.10} & 0 & {//} & 0 & -1.34 \\
        \hline\hline
    \end{tabular}
    \tablefoot{The energies are reported in eV. The values for the forsterite surface correspond to the most favorable Langmuir--Hinshelwood-type pathway, in which H$_2$CO is chemisorbed and NH$_2$ is initially physisorbed nearby. The remaining surface pathways are reported in Appendix~\ref{Appendix_A}. vdW, Int., and Prod. denote the vdW complex, intermediate, and product, respectively. TS1 and TS2 denote the first and second TSs along the corresponding pathway, respectively. A value of zero indicates that no barrier was resolved, whereas ``//'' indicates that no distinct stationary intermediate was located along that pathway. All stationary points were fully optimized at the indicated level of theory.}
    \label{tab:Formamide_formation}
\end{table*}

\section{{Discussion} and astrophysical implications}
\label{Astrophysical Implications}

The RNA world hypothesis suggests that nucleobases could form from HCN in water under early Earth conditions \citep{Leslie:2004}. However, HCN is very volatile and unlikely to have existed in the concentrations needed on early Earth, making it an improbable sole source for all the nucleobases required for life \citep{Saladino2006, Adande:2013,SALADINO201284}.

Formamide, on the other hand, has been proposed as a promising alternative to HCN in RNA formation \citep{Adande:2013} due to its low volatility. 
Formamide has been observed in several regions \citep{Solomon:1973,Rubin_1971,Lopez_2019,Codella_2017,Adande:2013}, including the centers of galactic sources such as Sagittarius SgrA and SgrB2 \citep{Gottlieb:1973}, and in comets such as C/1995 O1 (Hale-Bopp) and C/1996 B2 (Hyakutake; \citealt{Bockel:2000,Lis1997}). Furthermore, formamide has been hypothesized to exist on Jupiter, Saturn's moons, and other planets \citep{PARNELL2006322,Borucki:2002,LEVY2000609}. On grains, formamide has been observed around young stellar objects \citep{Schutte:1999}.
{However, the mechanisms underlying its formation in interstellar space and protoplanetary environments remain subjects of ongoing debate within the scientific community} \citep{Codella_2017, Barone_2015, Douglas_2022}. 
{Although observations have been interpreted as supporting gas-phase formamide formation \citep{Codella_2017}, theoretical studies indicate that the direct gas-phase NH$_2$ + H$_2$CO route remains chemically challenging because of competing bimolecular processes and the large barrier associated with H abstraction \citep{Douglas_2022,Rimola_2023}.}

{Hydrated Mg-bearing motifs may form when water interacts with Mg-rich silicate material under transient heating and cooling conditions. Shock-related processes provide a possible environment in which such hydration chemistry may occur.
Accretion shocks, spiral density waves, and bow shocks can locally increase temperatures and densities and may be accompanied by transient increases in water vapor and OH radicals, as suggested by \textit{Spitzer} and JWST observations of rotational H$_2$O and OH lines in sources such as NGC 1333 IRAS 4B \citep{Waterline:2012,vanGelder:2024}. Nebular shock models also indicate that such events can trigger ice sublimation, temporarily increase the partial pressure of H$_2$O, and affect cooling timescales \citep{DESCH:2002}. Under water-rich conditions, simulations show that H$_2$O adsorption on olivine surfaces can promote surface hydroxylation and the formation of amorphous brucite-like Mg(OH)$_2$ domains, with phyllosilicate phases forming over longer timescales or repeated thermal cycles \citep{Asaduzzaman:2015}. In addition, growth experiments in thin water films show that brucite nanosheets can be stabilized by intercalated molecular water layers, producing metastable hydrated structures with effective stoichiometry Mg(OH)$_2$·H$_2$O \citep{Langmuir:2023}. Since brucite dehydroxylation occurs only at temperatures above $\sim$350–400$^\circ$C \citep{Ball_Taylor_1961}, hydrated Mg-bearing motifs may persist after cooling in localized post-shock environments. These considerations motivate the use of Mg(OH)$_2\cdot$H$_2$O as a chemically defined hydrated Mg center to isolate the role of Mg coordination and proton transfer in the NH$_2$ + H$_2$CO reaction. In this way, the calculations provide elementary reaction energies and barriers that can be used to assess how hydrated Mg-bearing environments, if present locally, would affect formamide formation in future astrochemical models. Such chemically defined hydrated Mg-bearing species may also provide useful targets for future spectroscopic characterization and observational searches aimed at assessing the role of hydrated Mg chemistry in warm dust environments.}

{Evidence from the {Origins, Spectral Interpretation, Resource Identification, and Security--Regolith Explorer (OSIRIS-REx)} sample of asteroid Bennu, where formaldehyde, amines, amino acids, and anhydrous olivine were detected \citep{Glavin2025}, further supports the broader relevance of formaldehyde- and amine-bearing chemistry in primitive planetary material, while also illustrating that such chemistry can coexist with Mg-rich silicate phases.} {In the chemical evolution preceding the disk phase in the model of \citet{Quenard_2018}, gas-phase NH$_2$ undergoes transient enhancements during the late pre-stellar and warm-up stages, associated with nonthermal and thermal desorption, respectively. The model adopts binding energies of 3200~K for NH$_2$ and 4500~K for H$_2$CO, taken from \citet{WAKELAM201722}, which are consistent with NH$_2$ desorbing more readily than H$_2$CO. The full-disk model of \citet{Walsh:2014} predicts grain-surface H$_2$CO over a broad radial range, although the size of this surface reservoir varies strongly with distance from the central star. These model results provide physical motivation for considering reaction limits involving both gas-phase and adsorbed NH$_2$ reacting with surface-bound H$_2$CO. However, the binding energies employed in these astrochemical models are not specific to the mineral substrates considered here and therefore do not capture variations in adsorption strength arising from surface composition and local binding sites, which may affect the predicted balance between adsorption and desorption. } {The corresponding binding energies obtained here for H$_2$CO and NH$_2^{\bullet}$ considered separately, i.e., in the absence of the other reactant, are 0.84 and 1.08~eV, respectively, for Mg(OH)$_2\cdot$H$_2$O, and 0.90 and 1.20~eV, respectively, for the forsterite (010) surface (Table~\ref{tab:binding_energies}). For both Mg-bearing model systems, NH$_2^{\bullet}$ has a higher calculated binding energy than H$_2$CO, opposite to the relative ordering of the binding energies adopted for water-ice surfaces in the astrochemical models discussed above, highlighting the dependence of adsorption strength on the nature of the substrate. Although the relatively high binding energy of NH$_2^{\bullet}$ may favor its retention on the bare mineral surface, its actual surface abundance will also depend on its formation, accretion, diffusion, desorption, and competing reaction pathways. A quantitative estimate of this abundance will therefore require future dedicated astrochemical modeling studies.}

{At temperatures below 100 K, silicate surfaces may be covered by a water-ice layer, limiting direct access to the bare-mineral pathways examined here. Atomic H is an important competitor because it can hydrogenate H$_2$CO to form methanol \citep{Tielens_2013}. The calculated vibrationally adiabatic barrier for H addition is 0.20 eV \citep{Goumans_2011_forma}, which is comparable to the 0.25 eV C--N bond-forming barrier obtained for the isolated Mg(OH)$_2\cdot$H$_2$O model (Table \ref{tab:Formamide_formation}). This comparison does not apply to the most favorable Langmuir--Hinshelwood-type pathway on forsterite, for which C--N bond formation proceeds without a resolved barrier. However, competition between H and NH$_2$ addition cannot be determined from barrier heights alone, because it also depends on reactant abundances, adsorption and desorption, surface diffusion, and tunneling. The computed rate constants for the addition of H to H$_2$CO on grains range between 10$^2$ and 10$^3$ s$^{-1}$, and between 10$^5$ and 10$^7$ s$^{-1}$ in the gas phase, over the temperature range 60--100 K \citep{Song_2017}. At higher temperatures, between 299 and 603 K, experimental unimolecular rate constants for H$_2$CO hydrogenation range from 1.38 to 804.49 s$^{-1}$ \citep{Oehlers:2000}.}

\begin{figure*}
\centering
\includegraphics[width=\hsize]{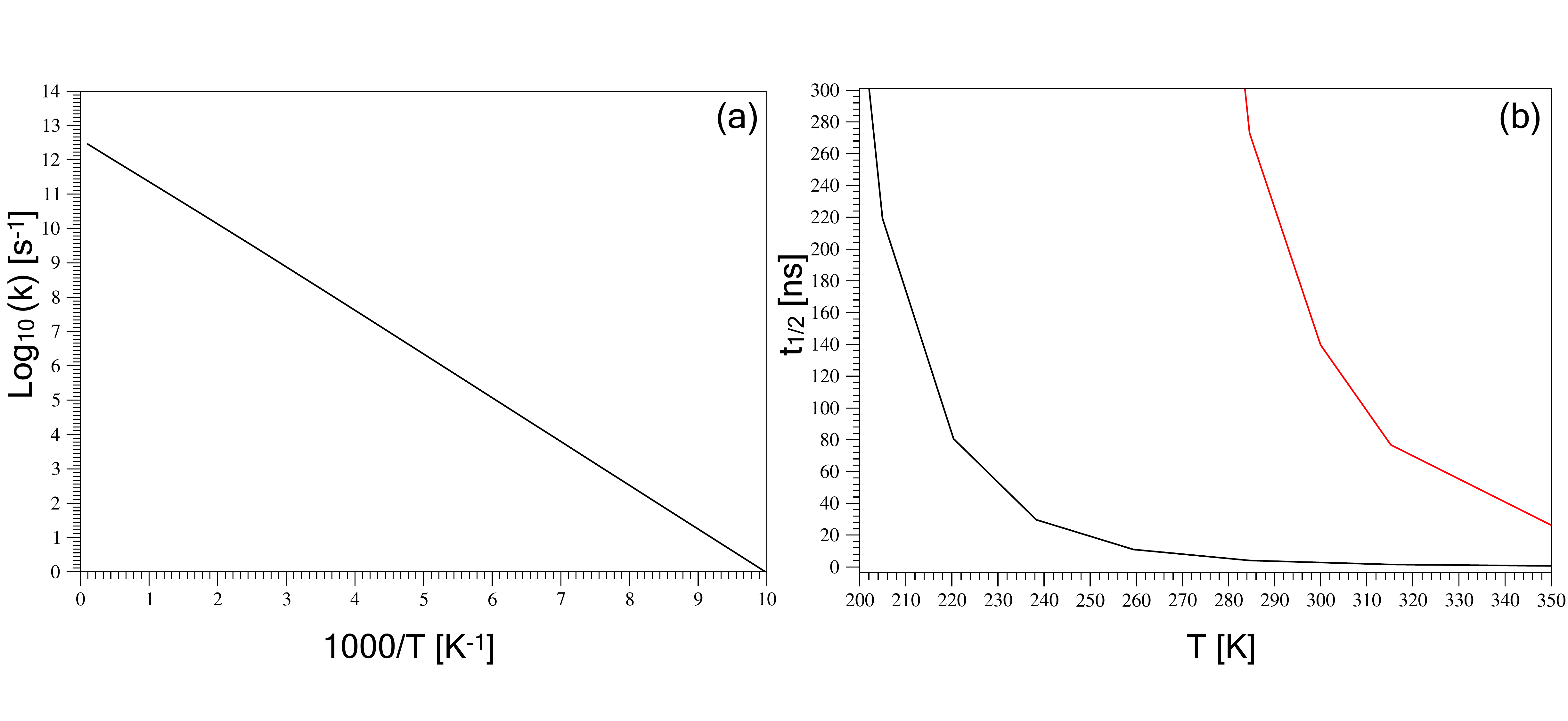}
\caption{{\textit{Panel a}: Arrhenius plot of log$_{10}(k)$ as a function of $1000/T$, where $T$ is the temperature in kelvin (K), for NH$_2$ addition to formaldehyde coordinated to the isolated Mg(OH)$_2$·H$_2$O molecular model at the M06-2X/pcseg-2 level of theory. The reported unimolecular rate constants correspond to classical rate constants corrected for zero-point harmonic vibrational energies. \textit{Panel b}: Half-lives ($t_{1/2}$), in nanoseconds (ns), as a function of temperature for forward evolution from the vdW complex toward the intermediate (black curve) and for NH$_2$ desorption from the vdW complex (red curve). The desorption half-lives were calculated from the estimated thermal desorption rate constants using $t_{1/2}=\ln 2/k_{\mathrm{des}}$ (see Appendix C). Numerical values are provided in Tables \ref{tab:fitted_rates}, \ref{tab:NH2_desorption_rates}, and \ref{tab:half-life}.}}
\label{fgr:rate_half_life}
\end{figure*}

{Unimolecular rate constants including quantized vibrational contributions for NH$_2$ addition to formaldehyde in the Mg(OH)$_2$·H$_2$O model are shown in Fig. \ref{fgr:rate_half_life}a. The logarithm of the rate constant varies linearly with the inverse temperature ($1000/T$), indicating Arrhenius-like behavior. Figure \ref{fgr:rate_half_life}b shows the corresponding half-lives ($t_{1/2}=\ln 2/k$) for forward evolution from the vdW complex toward the intermediate (black curve) and for NH$_2$ desorption from the vdW complex (red curve). The desorption half-lives were obtained from the estimated thermal desorption rate constants reported in Appendix C using $t_{1/2}=\ln 2/k_{\mathrm{des}}$. Over the reported temperature range, the characteristic time for forward evolution is shorter than that for NH$_2$ desorption. At 300 K, the forward half-life is approximately 2.6 ns, whereas the NH$_2$ desorption half-life is approximately 140 ns, corresponding to a forward rate approximately 54 times larger than the desorption rate.} {These values provide an estimate of the competition between forward reaction and desorption. Their quantitative interpretation remains subject to significant uncertainty because the desorption pre-exponential factor may be orders of magnitude larger when the rotational partition function of gas-phase NH$_2^{\bullet}$ is taken into account, while the binding energy derived from the isolated Mg(OH)$_2\cdot$H$_2$O model may be underestimated. These uncertainties may substantially affect the predicted competition between reaction and desorption.}

{The computed C--N bond-forming step shows thermally activated behavior up to approximately 300 K, with its rate primarily governed by the entrance barrier over this temperature range. Above this temperature, the forward-reaction half-life varies only weakly with temperature over the investigated range. Thus, once the NH$_2$--H$_2$CO--Mg(OH)$_2\cdot$H$_2$O encounter complex is formed, the computed elementary step is fast on the timescale of NH$_2$ desorption from the vdW complex. } {The calculations presented here provide substrate-specific molecular-level energetics that could guide future gas--grain astrochemical models assessing the contribution of these pathways under disk conditions.}

\section{Conclusions}
\label{Conclusions}

{This study presents quantum-chemical data for a potential pathway to interstellar formamide formation through the reaction of NH$_2$ with H$_2$CO in two Mg-bearing model systems chosen to probe chemistry in warm Mg-rich environments: the isolated Mg(OH)$_2\cdot$H$_2$O molecular model and the periodic forsterite (010) surface. I first benchmarked the selected theoretical methods for describing the PES of NH$_2$ addition to H$_2$CO in the presence of Mg(OH)$_2\cdot$H$_2$O. The benchmark shows that M06-2X/pcseg-1 and PBE+U5/DZP provide the closest agreement with the CCSD(T)-F12 reference among the tested DFT approaches.}

{In the Mg(OH)$_2\cdot$H$_2$O model, the reaction proceeds through the formation of a vdW complex, followed by the addition of NH$_2$ to formaldehyde to form a C--N bond and, after overcoming a small rotational barrier, by proton transfer to an O atom of Mg(OH)$_2\cdot$H$_2$O, forming a second water molecule coordinated to the Mg center. Compared with the corresponding bare gas-phase process, Mg(OH)$_2\cdot$H$_2$O promotes proton transfer, whereas H abstraction leading to formamide in the gas phase is associated with a large barrier.}

{On the forsterite (010) surface, the most favorable pathway occurs when NH$_2$ is weakly physisorbed near chemisorbed H$_2$CO. From this pre-reactive configuration, C--N bond formation and the subsequent surface-mediated proton-transfer step proceed without resolved barriers, yielding chemisorbed formamide. In contrast, direct attack by gas-phase NH$_2$ and the pathway starting from NH$_2$ chemisorbed on a neighboring Mg site require substantial activation barriers. Thus, forsterite promotes formamide formation through a local Langmuir--Hinshelwood-type pathway in which NH$_2$ is weakly physisorbed in close proximity to chemisorbed H$_2$CO.}

{For the Mg(OH)$_2\cdot$H$_2$O model, the unimolecular rate calculations show that, once the vdW complex is formed, conversion to the C--N-bonded intermediate is rapid and occurs on nanosecond timescales at temperatures approaching 300 K. Comparison with the estimated thermal desorption rate of NH$_2$ from the vdW complex shows that, at 300 K, the forward rate constant is approximately 54 times larger than the desorption rate constant.}

{Taken together, these results broaden the astrochemical picture of formamide formation beyond gas-phase and ice-mediated chemistry by identifying possible additional routes to its formation in Mg-bearing molecular and mineral environments. The calculated reaction pathways and unimolecular kinetics provide molecular-level information for future assessments of the potential contribution of these Mg-assisted routes under specific local conditions in warm Mg-rich environments. Formamide is observed in star-forming regions and comets, and identifying its formation pathways helps clarify how prebiotic material is chemically processed during the evolution from interstellar clouds to planetary systems. Formamide has the simplest peptide-like C(O)--N motif and has been proposed as a precursor in pathways toward peptide-bond formation and RNA-related molecular complexity \citep{SALADINO201284,Lopez_2019}. Establishing which astrophysical environments can support formamide formation is therefore important for linking its presence in star-forming regions and comets to the chemical evolution of material incorporated into planetary systems. By identifying potential Mg-assisted mechanisms, this study motivates future observational searches for relevant Mg-bearing species in warm Mg-rich environments and astrochemical modeling aimed at assessing the possible contribution of these pathways to interstellar formamide formation.}

\section*{{Data availability}}
All the data, for example input, output, and Cartesian coordinate files, from the present work are available in the Zenodo repository {(https://doi.org/10.5281/zenodo.14907634)}.

\begin{acknowledgements}
This project has received funding from the AIAS-AUFF Fellowship program funded by Aarhus University Research Foundation and Aarhus Institute of Advanced Studies (AIAS) at Aarhus University. This work has been supported by the Danish National Research Foundation through the Center of Excellence “InterCat” (Grant agreement no.: DNRF150). 
This work was supported by DeiC National HPC (g.a. DeiC-AU-N5-2026172). I acknowledge the state of Baden-Württemberg through bwHPC and the German Research Foundation (DFG) through grant no INST 40/575-1 FUGG (JUSTUS 2 cluster).  I thank Johannes Kästner for computational resource support and Raffaele Cheula for providing Sella input scripts.
I thank the University of Perugia for its research funding support under the assegno di ricerca scheme. I acknowledge the Alexander von Humboldt Foundation for postdoctoral fellowship support.

\end{acknowledgements}

\bibliographystyle{aa} 
\bibliography{aa} 

\begin{appendix}
\section{Formamide formation on the forsterite (010) surface} \label{Appendix_A}

{For this work, I constructed a 4 x 3 x 1 periodic model of the forsterite (010) surface, which has been identified as the most stable surface in previous studies \citep{deLeeuw:2001, Bruno2014, Zamirri:2017, Campisi_2022, Campisi_MNRAS_2024} and confirmed by the broader scientific community \citep{Ruiz:2014, Ruiz:2015, Zamirri:2017, Garcia-Gil:2013}. This surface is of particular importance in interstellar space, as it facilitates the adsorption of H, CO, PAHs, and many other species, with binding energies ranging from 0.5 eV to several eV.} Consequently, its polar surface is highly reactive and capable of hosting atomic and molecular species.

The (010) surface of the forsterite model, which I used to study the reaction of formaldehyde with the NH$_2$ radical, has been characterized in a prior work \citep{Campisi_JPCA_2021,Campisi_2022}. It is formed by a 4 x 3 x 1 slab model with an orthorhombic structure, space group Pbnm, and has unit cell dimensions of 19.00 x 17.94 x 35.20~\AA, with angles $\alpha$, $\beta$, and $\gamma$ of 90$^\circ$. Atomistically, the surface is characterized by under-coordinated Mg atoms, each binding to only three neighboring O atoms {and, therefore, being able to host organic molecules to complete their coordination.}

\begin{figure}
\centering
  \includegraphics[width=\hsize]{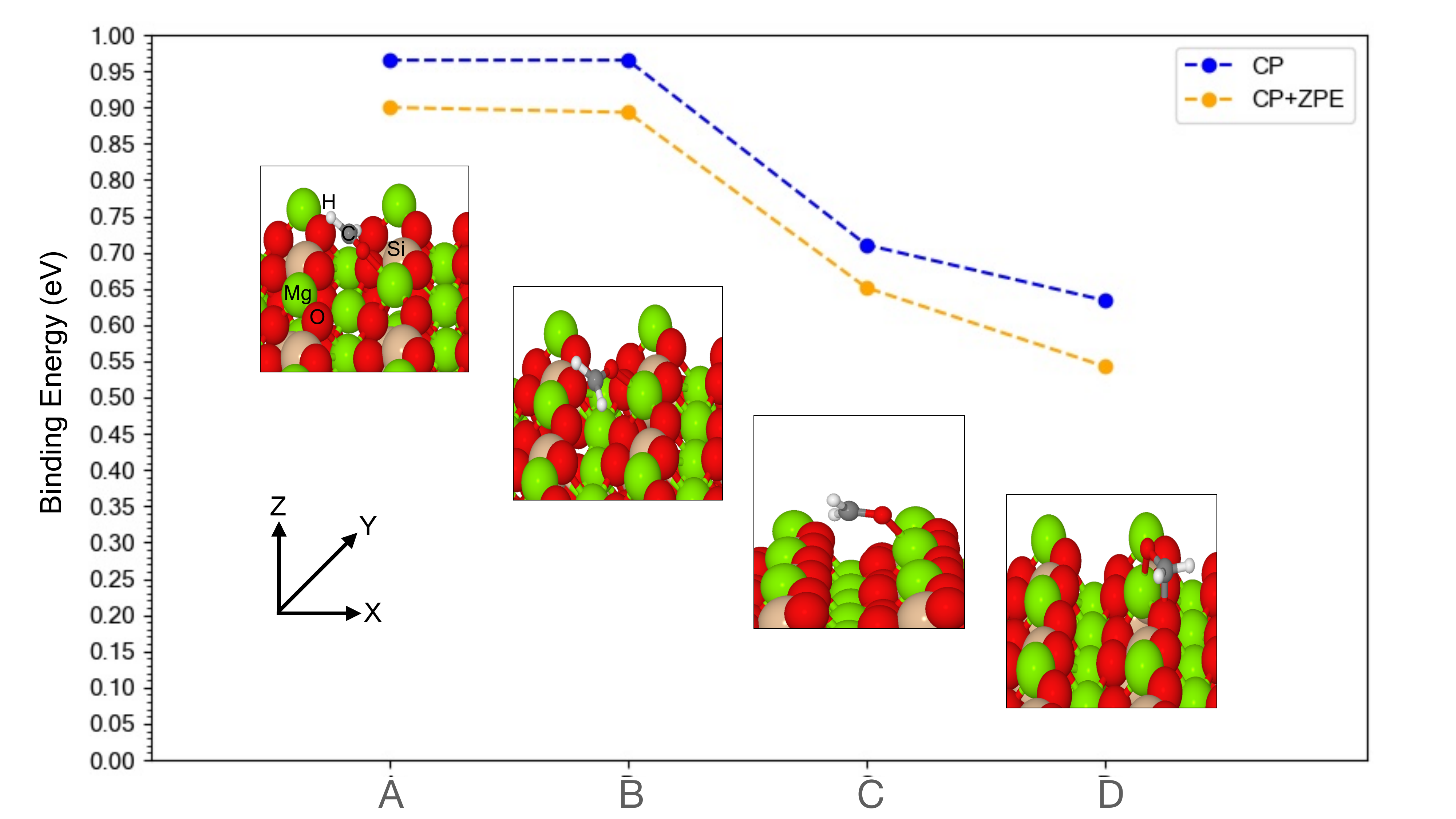}
  \caption{Binding energies of adsorption configurations, A through D, of formaldehyde on the (010) surface of forsterite shown at the PBE/DZP level. The counterpoise-corrected (CP) binding energies are indicated by blue dots, while both counterpoise and zero-point-corrected binding energies (CP+ZPE) are indicated by orange dots. Atomic labels are provided for the atoms in configuration A as an example, and these labels apply to all the structures shown. In the diagrams, Mg atoms are represented by green balls, O atoms by red balls, C atoms by gray balls, Si atoms by beige balls, and H atoms by white balls. All configurations, A to D, are oriented with the surface facing the vacuum region along the Z axis.}
  \label{fgr:forma_ads}
\end{figure}

To study the formation of formamide, I first identified different binding conformations of formaldehyde on (010) forsterite, as reported in Fig. \ref{fgr:forma_ads}. I found four possible adsorption sites (A, B, C, and D; see Fig. \ref{fgr:forma_ads}). Adsorption sites A and B are similar, with equal binding energies, where the molecule chemisorbs to an O atom of the Mg center, resulting in a tilt with respect to the plane of the surface and forming a hydrogen bond with the coordinated O atom of the Mg center. The binding energies for A and B are higher, indicating more stable adsorption sites. On the other hand, adsorption site C results in a parallel adsorption with respect to the $x$-axis of the surface, while configuration D, the least stable one, results in the formation of O-Mg and C-Mg bonds at the Mg-O center. The binding energies of the identified configurations range from {0.5 to 1.0 eV}, in agreement with theoretical studies \citep{Mates-Torres:ui5001, VINOGRADOFF2024118558}. 

\begin{table}
    \centering
\caption{Binding energies (BE) of the Mg(OH)$_2\cdot$H$_2$O and forsterite (010) surface reaction pathways.}
    \begin{tabular}{@{}lcc@{}}
        \hline\hline
     {Fragments} & 
     {BE$_{\mathrm{CP}}$} & 
     {BE$_{\mathrm{CP+ZPE}}$} \\
        \hline

        {H$_2$CO + Mg(OH)$_2\cdot$H$_2$O} &
        {0.95} &
        {0.84} \\

        {NH$_2^{\bullet}$ + Mg(OH)$_2\cdot$H$_2$O} & {1.21}  & {1.08} \\

        {NH$_2^{\bullet}$ + H$_2$CO--Mg(OH)$_2\cdot$H$_2$O} {(vdW)} &
        {0.46} &
        {0.35} \\

        {NH$_2$CH$_2$O$^{\bullet}$ + Mg(OH)$_2\cdot$H$_2$O} &
        {0.95} &
        {0.82} \\

        {NH$_2$CHO + [Mg(OH)(H$_2$O)$_2$]$^{\bullet}$} &
        {0.48} &
        {0.48} \\

        \hline\hline
        {H$_2$CO + forsterite} &
         {0.97} &
         {0.90} \\
        {NH$_2^{\bullet}$ + forsterite} & {1.32}  & {1.20} \\

        {NH$_2^{\bullet}$ + H$_2$CO--forsterite {(Int. 2)}} &
        {1.35} &
        {1.23} \\

        {NH$_2^{\bullet}$ + H$_2$CO--forsterite (vdW)} &
        {0.15} &
        {0.10} \\

        {NH$_2$CH$_2$O$^{\bullet}$ + forsterite} &
        {1.33} &
        {1.25} \\

        {NH$_2$CHO + [H--forsterite]$^{\bullet}$} &
        {1.27} &
        {0.98} \\

        \hline\hline
    \end{tabular}
    \tablefoot{Binding energies are reported in eV. BE$_{\mathrm{CP}}$ denotes the counterpoise-corrected binding energy, whereas BE$_{\mathrm{CP+ZPE}}$ includes both counterpoise and zero-point-energy corrections. Each entry in the first column lists the potentially desorbing species and the remaining Mg-bearing complex or forsterite surface as separate fragments. For H$_2$CO + Mg(OH)$_2\cdot$H$_2$O, NH$_2^{\bullet}$ + Mg(OH)$_2\cdot$H$_2$O, H$_2$CO + forsterite, and NH$_2^{\bullet}$ + forsterite, the reported binding energies correspond to the isolated species bound to the respective Mg-bearing model or forsterite (010) surface in the absence of the other reactant. The entries labeled vdW and Int.~2 refer to desorption from the corresponding configurations along the reaction pathways shown in Figs.~3 and~\ref{fgr:periodic_PES}.}
    \label{tab:binding_energies}
\end{table}

\begin{figure*}
\centering
\includegraphics[width=\hsize]{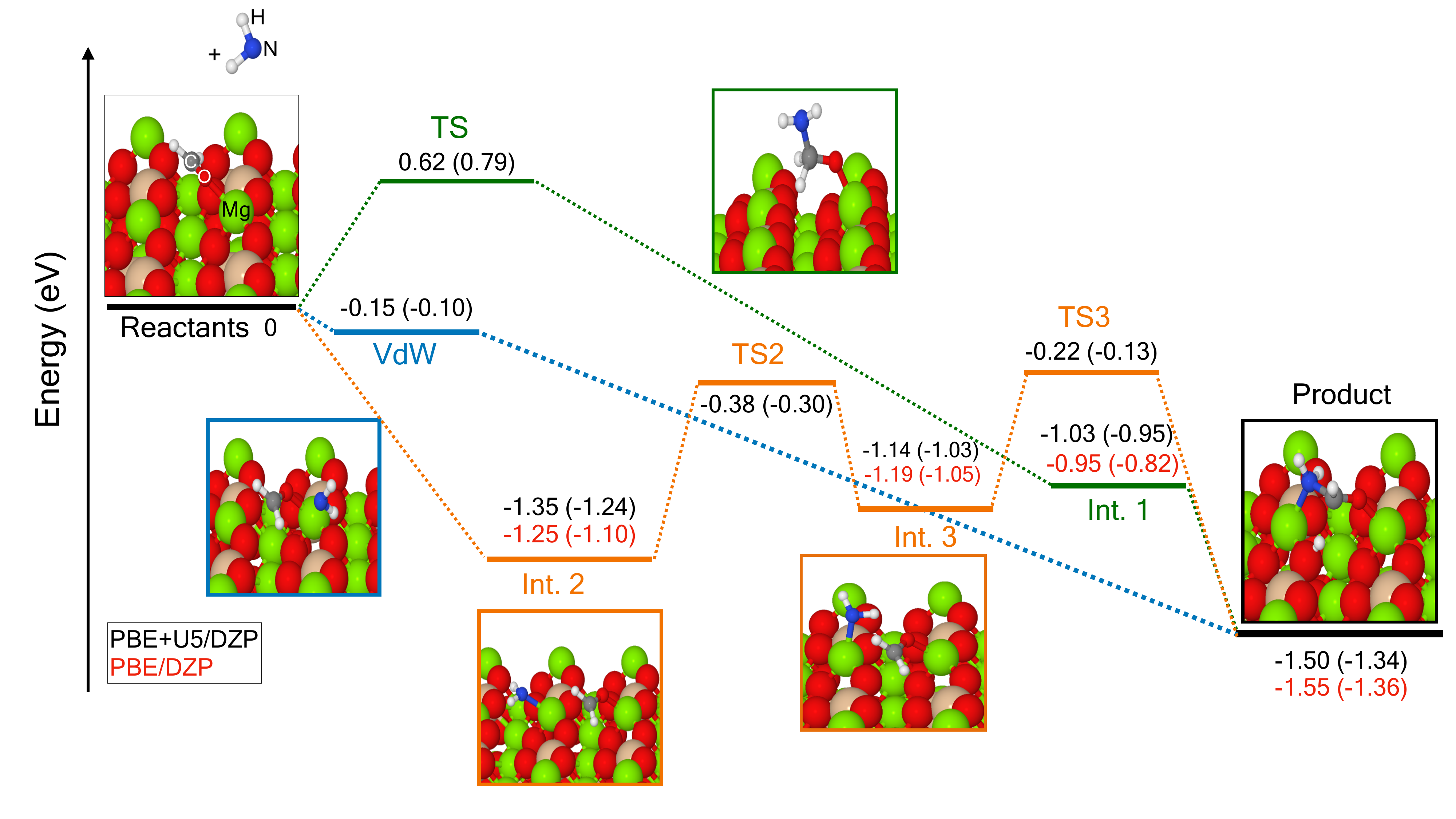}
\caption{{{Schematic representation of the} PES for the reaction between NH$_2$ and formaldehyde on the (010) forsterite surface. {Ball-and-stick models of the reactants, the vdW complex, intermediates (Int.) 1--3, and the product are shown above the PES, with corresponding atomic labels provided as an example for the reactant structures and applicable to the remaining geometrical structures.}
The values, in eV, obtained at the PBE+U5/DZP level are reported above the corresponding values at the PBE/DZP level, {where available}. {All reported values include the counterpoise correction, whereas values in parentheses additionally include the zero-point-energy correction. The green, blue, and orange profiles represent the Eley--Rideal-like pathway, the local Langmuir--Hinshelwood-type pathway starting from the weakly physisorbed vdW complex, and the Langmuir--Hinshelwood-type pathway starting from NH$_2$ chemisorbed on a neighboring Mg site, respectively.}}}
\label{fgr:periodic_PES}
\end{figure*}

{In Fig. \ref{fgr:periodic_PES}, I report the PES for three pathways for the addition of NH$_2$ to formaldehyde on the forsterite (010) surface. In the Eley--Rideal-like pathway, gas-phase NH$_2$ directly attacks the C atom of chemisorbed formaldehyde. At the PBE+U5/DZP level, C--N bond formation requires overcoming a counterpoise-corrected barrier of 0.62 eV, which increases to 0.79 eV after the zero-point-energy correction, and leads to the intermediate Int.~1.}

{An alternative Langmuir--Hinshelwood-type pathway, shown in orange in Fig. \ref{fgr:periodic_PES}, begins with the adsorption of NH$_2$ on a neighboring surface Mg atom without a resolved barrier, forming Int.~2. Reorientation of NH$_2$ toward the reaction center requires overcoming a zero-point-energy-corrected rotational barrier of 0.94 eV, leading to Int.~3. Subsequent C--N bond formation requires overcoming an additional barrier of 0.90 eV and leads to the product.}

{Finally, when NH$_2$ is weakly physisorbed near chemisorbed formaldehyde, forming the vdW complex shown in Fig. \ref{fgr:periodic_PES}, C--N bond formation and the subsequent proton transfer to a neighboring surface O atom proceed without resolved barriers, as confirmed by the NEB calculations reported in Appendix B, yielding a more stable chemisorbed formamide product. Although diffusive processes are certainly interesting and deserve further investigation, they are outside the scope of this work, and I will address them more thoroughly in a separate study.}

{In Fig. \ref{fgr:periodic_PES}, the effect of the zero-point correction is only to reduce the exothermicity of the reaction, with differences from the uncorrected values ranging between 0.1 and 0.2 eV. {The binding energies of the relevant species involved in the reaction pathways, including the formamide product, are reported in Table \ref{tab:binding_energies}.}}

\section{NH$_2$ addition and proton transfer mechanism} \label{Appendix_B}

\begin{figure}
\centering
    \includegraphics[width=\hsize]{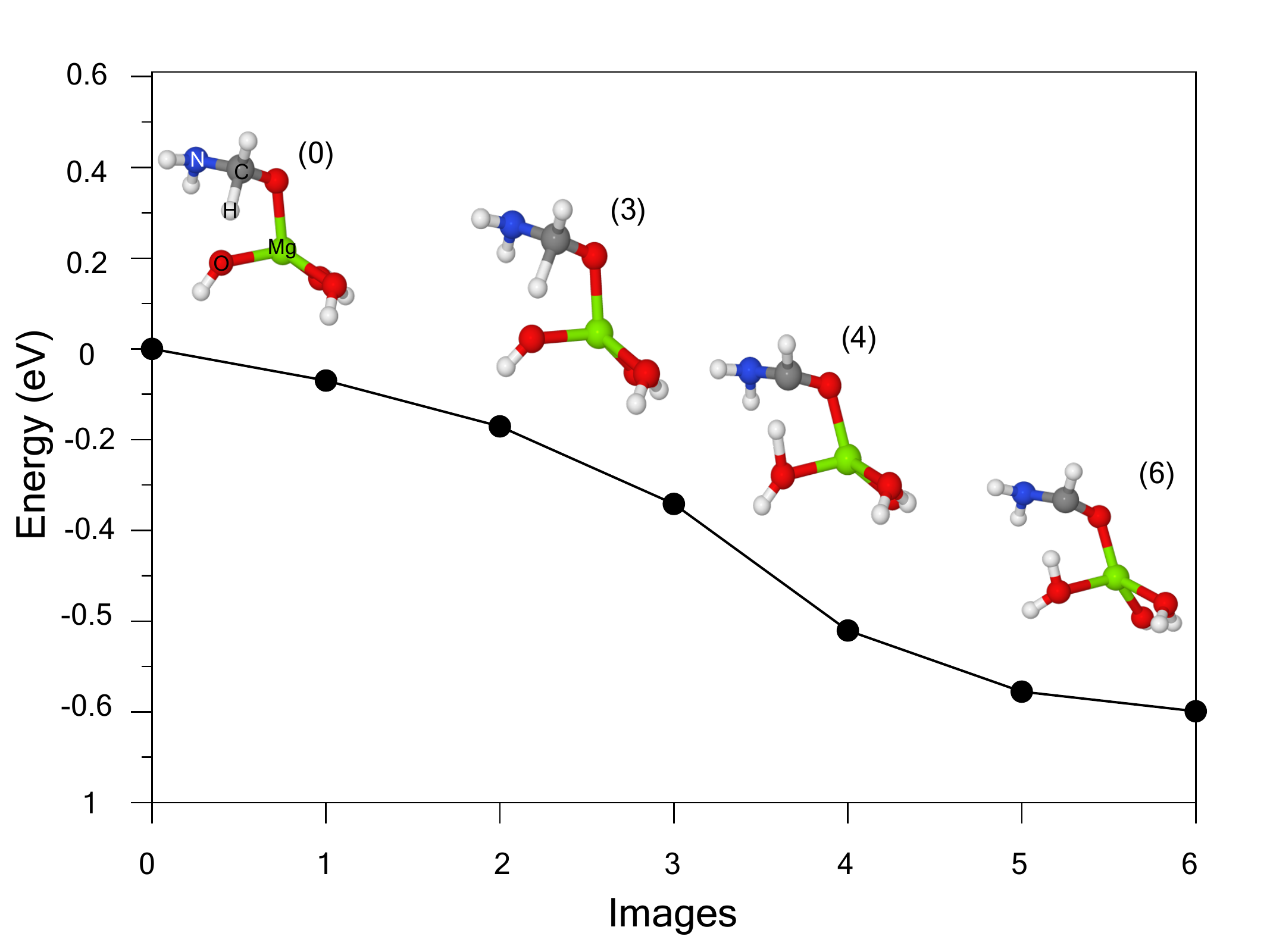}
  \caption{{CI-NEB profile for proton transfer following the second TS in the Mg(OH)$_2\cdot$H$_2$O pathway shown in Fig. \ref{fgr:PES}. The profile shows that proton transfer proceeds without an additional resolved barrier, leading to the Mg-coordinated formamide product. Relative electronic energies are reported in eV with respect to image 0. The $x$-axis indicates the image index. Optimized geometries for images 0, 3, 4, and 6 are shown along the reaction pathway. Atomic labels are provided for image 0 and apply to the other images. The CI-NEB calculation was performed at the M06-2X/pcseg-2 level of theory.}}
  \label{fgr:ProtonTransfer}
\end{figure}

{Figure \ref{fgr:ProtonTransfer} shows the CI-NEB calculation used to determine whether an additional barrier is present for proton transfer following the rotational barrier of the C--N-bonded intermediate in the Mg(OH)$_2\cdot$H$_2$O pathway. The figure provides a detailed view of the reaction pathway from the second TS to the product shown in Fig. \ref{fgr:PES}.}

In Fig. \ref{fgr:ProtonTransfer}, images 0 and 6 represent high-energy intermediates identified through an initial CI-NEB calculation, which interpolated 26 images from reactants to products (Fig. \ref{fgr:PES}). Subsequently, images 0 and 6 (as labeled in Fig. \ref{fgr:ProtonTransfer}) were refined by interpolating five additional images. {The refined profile shows that proton transfer from the C atom of the intermediate (image 0) to an O atom of Mg(OH)$_2\cdot$H$_2$O proceeds without an additional resolved barrier.}

\begin{figure}
\centering
\includegraphics[width=\hsize]{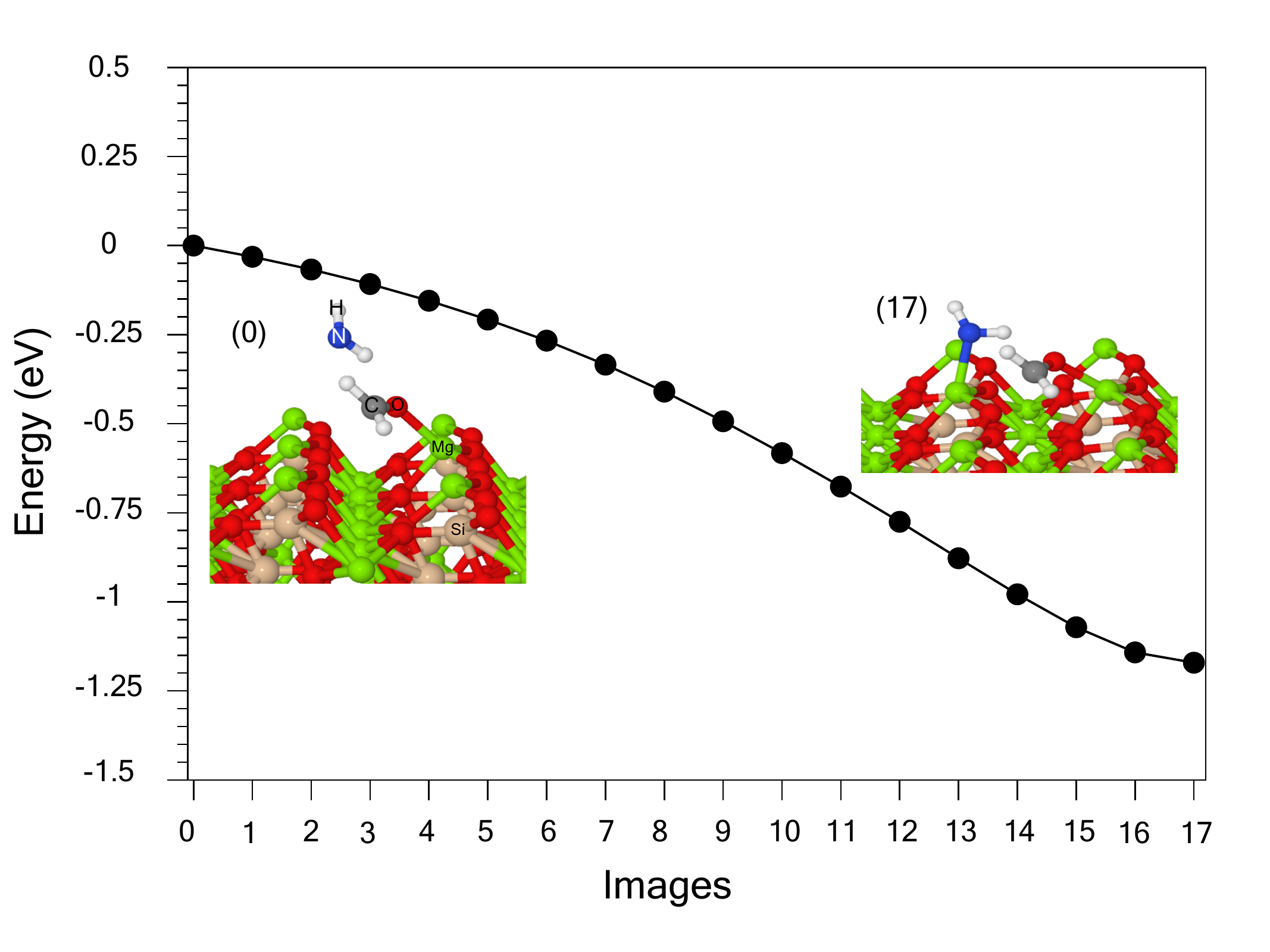}
\caption{{CI-NEB profile for the adsorption of NH$_2$ on a neighboring Mg site of the forsterite (010) surface. The profile shows that this adsorption step proceeds without a resolved barrier. Relative electronic energies are reported in eV with respect to image 0. The $x$-axis indicates the image index. Optimized geometries for images 0 and 17 are shown along the reaction pathway. Atomic labels are provided for image 0 and apply to the other images. The CI-NEB calculation was performed at the PBE+U5/DZP level of theory.}}
\label{fgr:N-Mg_NEB}
\end{figure}

\begin{figure}
\centering
\includegraphics[width=\hsize]{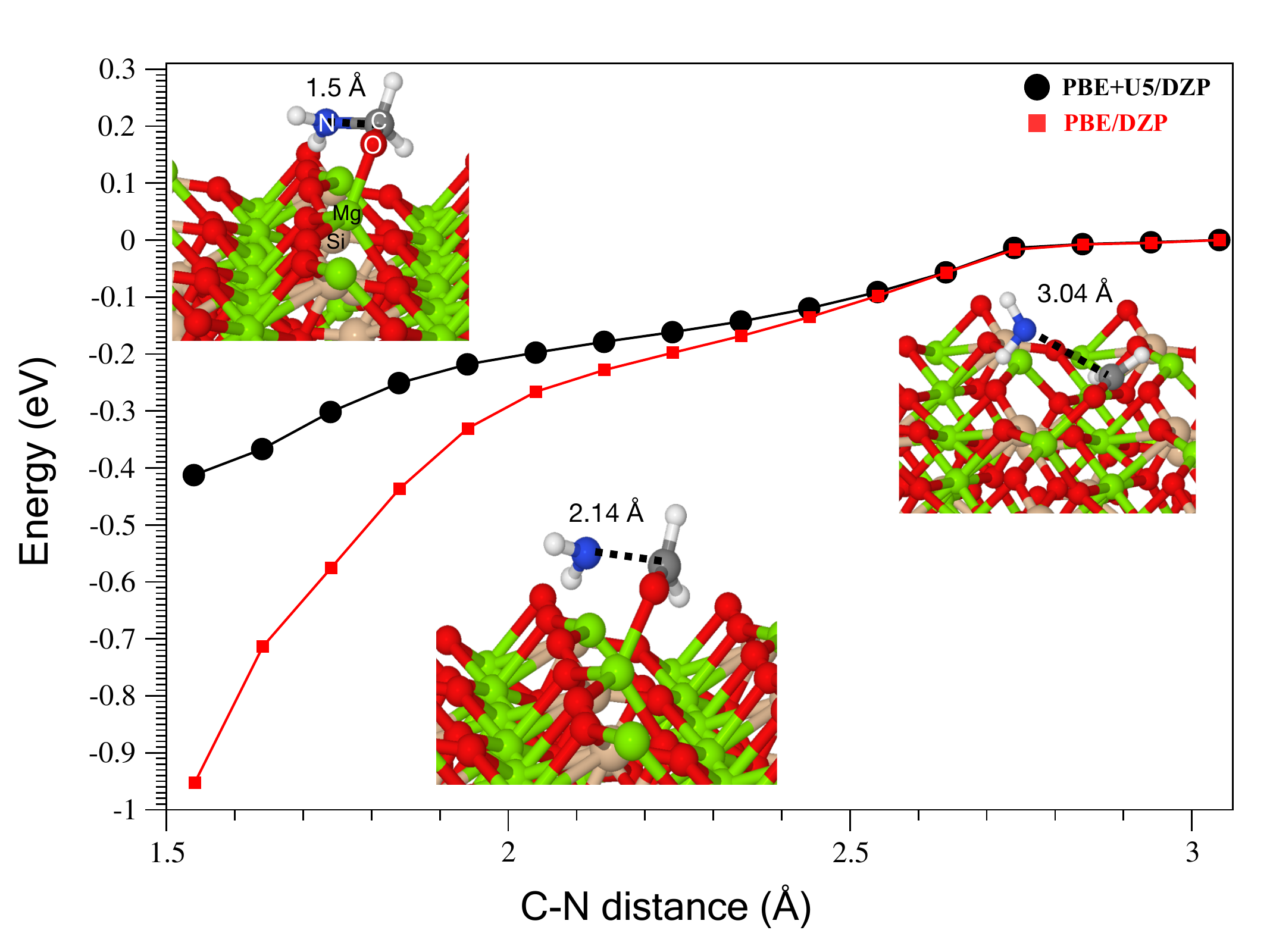}
\caption{Relaxed scan{s} of the C--N distance for the addition of weakly physisorbed NH$_2$ to H$_2$CO adsorbed on the forsterite (010) surface. {Circles and squares denote the PBE+U5/DZP and PBE/DZP profiles, respectively. Both scans show} that C--N bond formation proceeds without a resolved barrier. Relative electronic energies are reported in eV with respect to the
structure with a C--N distance of 3.04~$\AA$ {for each method}.
The $x$-axis shows the C--N distance in~$\AA$. Atomic labels are provided for the
first structure on the left and apply to all subsequent structures.}
\label{fgr:N-C_Scan}
\end{figure}

{For the reactions on the periodic forsterite (010) surface, Fig. \ref{fgr:N-Mg_NEB} shows the CI-NEB profile for the adsorption of NH$_2$ on a neighboring surface Mg atom, corresponding to the initial step of the orange pathway in Fig. \ref{fgr:periodic_PES}. No resolved barrier is found for this adsorption step. Figure \ref{fgr:N-C_Scan} reports the relaxed scan for C--N bond formation from the weakly physisorbed vdW complex. {Both PBE+U5/DZP and PBE/DZP profiles show a continuous decrease in energy as the C--N distance is shortened, with no intervening maximum along the reaction pathway. The two profiles remain similar at longer C--N distances and progressively separate as the C--N bond develops. The barrier-less nature of the C--N bond-forming step is therefore consistently observed at both the PBE+U5/DZP and PBE/DZP levels.}} 

\begin{figure}
\centering
  \includegraphics[width=\hsize]{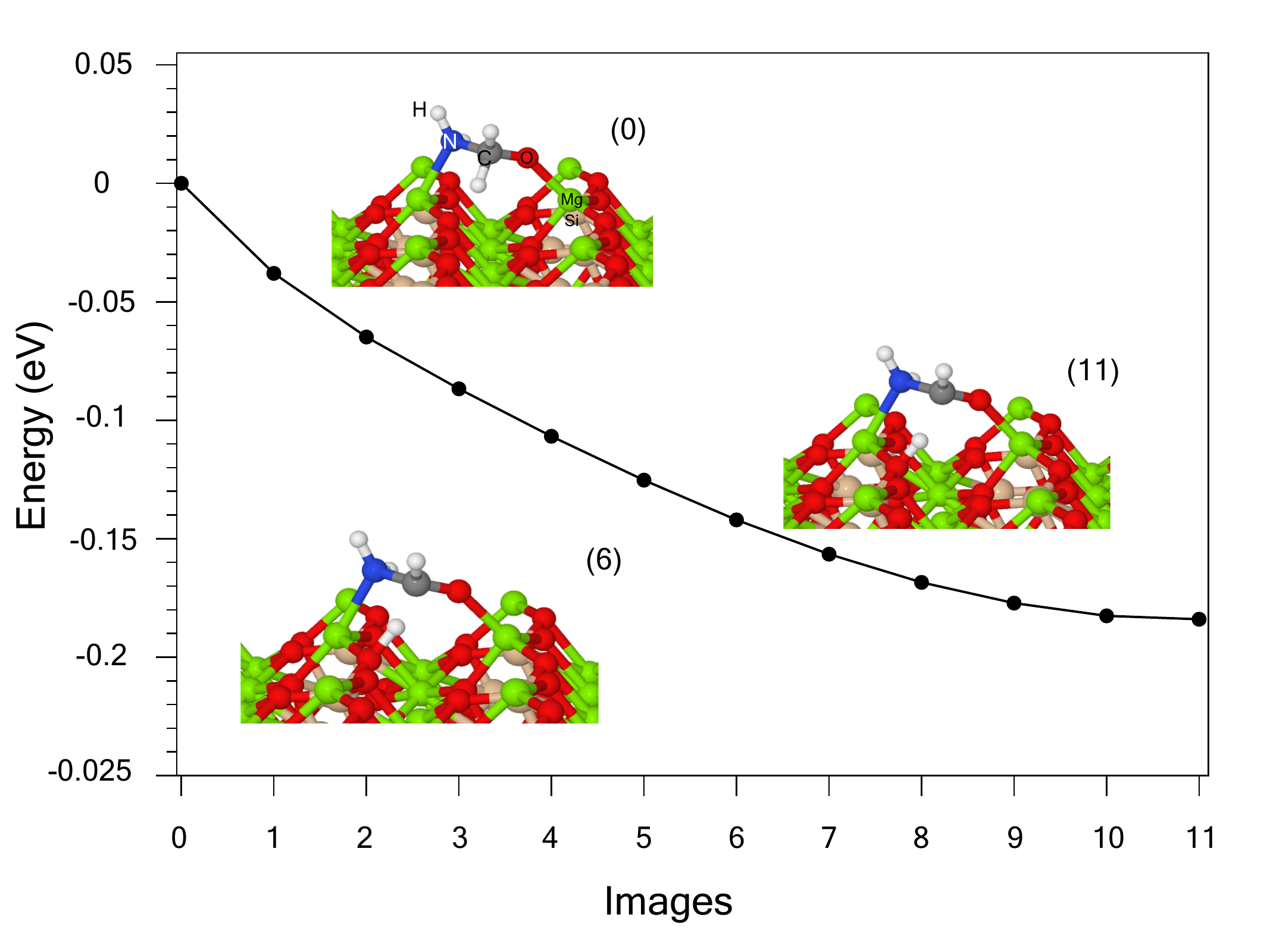}
  \caption{{CI-NEB profile for proton transfer from the C--N-bonded intermediate to a neighboring O atom on the forsterite (010) surface. The profile shows that proton transfer leading to chemisorbed formamide proceeds without an additional resolved barrier. Relative electronic energies are reported in eV with respect to image 0. The $x$-axis indicates the image index. Optimized geometries for images 0, 6, and 11 are shown along the reaction pathway. Atomic labels are provided for image 0 and apply to the other images. The CI-NEB calculation was performed at the PBE+U5/DZP level of theory.}}
  \label{fgr:ProtonTransfer_periodic}
\end{figure}

{Following the use of Voronoi charge analysis to investigate the electronic contribution to barrier-less C--H bond cleavage of PAHs on defective forsterite \citep{Campisi_2022}, Voronoi charges \citep{FonsecaGuerra_2004} were also considered here to investigate the origin of the barrier-less C--N bond formation starting from the pre-reactive vdW complex, in which NH$_2$ is weakly physisorbed close to H$_2$CO chemisorbed on forsterite. The resulting charges were compared with those obtained for the other Langmuir--Hinshelwood and Eley--Rideal pathways shown in Fig.~\ref{fgr:periodic_PES} and for the isolated Mg(OH)$_2\cdot$H$_2$O model (Table~\ref{tab:voronoi_charges}). Overall, no substantial differences in the N and C charges are observed among the configurations preceding C--N bond formation in the different pathways. This indicates that the initial physisorption of NH$_2$ in the vdW complex does not induce substantial charge transfer or strong electronic activation of the reactants. However, after C--N bond formation along the pathway starting from the vdW complex, the N charge becomes markedly less negative, approaching zero ($q_{\rm N}=-0.012\,e$). This behavior is not observed in the corresponding C--N-bonded configurations of the Eley--Rideal pathway, the Langmuir--Hinshelwood pathway in which NH$_2$ is initially chemisorbed on a neighboring Mg site, or the isolated Mg(OH)$_2\cdot$H$_2$O model. A strong dependence of surface reactivity on molecular orientation has also been reported previously for PAHs on defective forsterite, where parallel adsorption leads to chemisorption, whereas a perpendicular orientation promotes barrier-less C--H bond cleavage \citep{Campisi_2022}. When NH$_2$ is chemisorbed on a neighboring Mg site, its interaction with the surface restricts the orientations accessible to the radical, preventing direct access to the favorable geometry associated with barrier-less C--N bond formation; a reorientation step is therefore required before NH$_2$ addition.} {A more detailed assessment of the electronic and geometrical contributions underlying this effect is beyond the scope of the present work. Figure \ref{fgr:ProtonTransfer_periodic} shows the subsequent proton transfer to a neighboring surface O atom. No resolved barriers are found for either step, supporting the local Langmuir--Hinshelwood-type pathway shown in blue in Fig. \ref{fgr:periodic_PES}.}

\begin{table}
\centering
\caption{Voronoi charges of the N and C atoms in representative structures of the Mg(OH)$_2\cdot$H$_2$O and forsterite (010) surface reaction pathways.}
{
\begin{tabular}{lrr}
\hline
Structure & $q_{\rm N}$ & $q_{\rm C}$ \\
\hline

\multicolumn{3}{l}{Isolated model:} \\
\hline
vdW & -0.073 & 0.214 \\
Mg(OH)$_2\cdot$H$_2$O--NH$_2$CH$_2$O & -0.161 & 0.143 \\

\hline
\multicolumn{3}{l}{HL 1:} \\
\hline
vdW & -0.084 & 0.232 \\
forsterite--NH$_2$CH$_2$O & -0.012 & 0.160 \\

\hline
\multicolumn{3}{l}{HL 2:} \\
\hline
Int.~3 & -0.067 & 0.219 \\
Product  & -0.101 & 0.114 \\

\hline
\multicolumn{3}{l}{ER:} \\
\hline
NH$_2$ + H$_2$CO--forsterite & -0.075 & 0.221 \\
Int.~1 & -0.090 & 0.147 \\

\hline
\end{tabular}
}
\tablefoot{The Voronoi charges $q_{\rm N}$ and $q_{\rm C}$ are reported in units of $e$ for the N and C atoms, respectively. HL 1 and HL 2 denote the Langmuir--Hinshelwood pathways starting from weakly physisorbed NH$_2$ in the vdW complex and NH$_2$ chemisorbed on a neighboring Mg site, respectively. ER denotes the Eley--Rideal pathway involving gas-phase NH$_2$ approaching chemisorbed H$_2$CO. Int.~1, Int.~3, and Product refer to the corresponding structures in the PES shown in Fig.~\ref{fgr:periodic_PES}. All values were computed at the PBE/DZP level of theory.}
\label{tab:voronoi_charges}
\end{table}

\clearpage
\section{Rate constants} \label{Appendix_C}

{The computed rate constants for the forward reaction, {reported in Table~\ref{tab:QVK}}, were fitted using the modified Kooij--Arrhenius expression commonly adopted in astrochemical kinetic models \citep{UMIST}:}

\begin{eqnarray}
{k} & {=} & {\alpha \left( \frac{T}{300 \text{K}} \right)^{\beta} \exp \left( -\frac{\gamma}{T} \right),} \label{eq5}
\end{eqnarray}

{\noindent where $k$ is the rate constant in s$^{-1}$, $\alpha$ is the pre-exponential fitting parameter in s$^{-1}$, $\beta$ is the dimensionless temperature-dependence parameter, and $\gamma$ is the activation parameter expressed in K.}

{Table \ref{tab:parameters} reports the fitting parameters obtained with a Monte Carlo-based fitting code \citep{Campisi:2025}, employing the modified Kooij--Arrhenius expression reported in Eq.~\ref{eq5}. The code is available in the Zenodo repository associated with this article (see the "Data availability" section).} 

{The initial values of the fitting parameters $\alpha$ and $\gamma$ were sampled within the ranges $(1 \times 10^{-25},\,1 \times 10^{18})$~s$^{-1}$ and $(0.05,\,6000)$~K, respectively, to initialize the fitting procedure. The parameter $\beta$ was fixed to 0, corresponding to the classical Arrhenius limit of the Kooij--Arrhenius expression. A total of 5000 initial parameter pairs for $\alpha$ and $\gamma$ were selected through Monte Carlo sampling and subsequently optimized.}
 
{The fit yielded a root-mean-square error of $2.49 \times 10^{-2}$ for the $\log_{10}(k)$ values and an $R^2$ value of 0.99996. The DFT-computed rate constants and the corresponding values obtained from Eq.~\ref{eq5} are reported in Table~\ref{tab:fitted_rates}.}

{To assess the possible competition between forward evolution of the vdW complex toward the intermediate and desorption back to the separated reactants, the thermal desorption rate of NH$_2$ from the vdW complex was estimated. The desorption rate was calculated using a first-order thermal desorption expression,}

\begin{eqnarray}
{k_{\mathrm{des}}(T)} & {=} & {\nu \exp \left( -\frac{E_{\mathrm{des},K}}{T} \right),} \label{eq:desorption}
\end{eqnarray}

{\noindent where $E_{\mathrm{des},K}$ is the NH$_2$ desorption energy expressed in K. The desorption energy was taken as the M06-2X/pcseg-2 energy difference between the separated reactants and the vdW complex. The resulting zero-point-corrected binding energy at the M06-2X/pcseg-2 level of theory is $E_{\mathrm{des}} = 0.32$~eV, corresponding to $E_{\mathrm{des},K} = 3713.45$~K. The pre-exponential factor $\nu$ was estimated from the low-frequency vibrational mode of the vdW complex associated with the relative motion of NH$_2$ with respect to the remainder of the complex. This mode is located at 39.34~cm$^{-1}$, corresponding to $\nu = 1.18 \times 10^{12}$~s$^{-1}$.} {The estimated NH$_2$ desorption rates are reported in Table \ref{tab:NH2_desorption_rates}.}

\begin{table}
    \centering
\caption{Inverse temperature and unimolecular rate constants for NH$_2$ addition to H$_2$CO on Mg(OH)$_2\cdot$H$_2$O.}
    \begin{tabular}{cc}
        \hline
        \hline
        $1000/T$ & $\log_{10}(k)$ \\
        \hline
0.10 & 12.47 \\
0.44 & 12.05 \\
0.78 & 11.63 \\
1.12 & 11.21 \\
1.47 & 10.79 \\
1.81 & 10.37 \\
2.15 & 9.95 \\
2.49 & 9.52 \\
2.83 & 9.09 \\
3.17 & 8.67 \\
3.51 & 8.23 \\
3.86 & 7.80 \\
4.20 & 7.37 \\
4.54 & 6.93 \\
4.88 & 6.50 \\
5.22 & 6.07 \\
5.56 & 5.63 \\
5.90 & 5.19 \\
6.25 & 4.76 \\
6.59 & 4.32 \\
6.93 & 3.89 \\
7.27 & 3.45 \\
7.61 & 3.02 \\
7.95 & 2.58 \\
8.29 & 2.15 \\
8.63 & 1.71 \\
8.98 & 1.28 \\
9.32 & 0.84 \\
9.66 & 0.41 \\
10.00 & -0.02 \\
        \hline
        \hline
    \end{tabular}
\tablefoot{The inverse temperature is reported as $1000/T$, with $T$ in K. The rate constants $k$, in s$^{-1}$, are expressed as $\log_{10}(k)$ and refer to the entrance step in the isolated Mg(OH)$_2\cdot$H$_2$O molecular model at the M06-2X/pcseg-2 level of theory.}
    \label{tab:QVK}
\end{table}

\begin{table}
    \centering
\caption{Fitting parameters for the unimolecular rate constants of NH$_2$ addition to H$_2$CO on Mg(OH)$_2\cdot$H$_2$O.}
    \begin{tabular}{@{}ccc@{}}
        \hline\hline
          {$\alpha$} & {$\beta$} & {$\gamma$} \\
        \hline
        {4.272 x 10$^{12}$}  & {0.000} & {2.906 x 10$^{3}$} \\
        \hline\hline
    \end{tabular}
\tablefoot{The fitting parameters are $\alpha$ in s$^{-1}$, $\beta$ dimensionless, and $\gamma$ in K. They were obtained by fitting the rate constants for the entrance step in the isolated Mg(OH)$_2\cdot$H$_2$O molecular model. The rate constants were computed at the M06-2X/pcseg-2 level of theory and fitted using Eq.~\ref{eq5}, with $\beta$ fixed to 0.}
    \label{tab:parameters}
\end{table}

\begin{table}
    \centering
\caption{Temperature, DFT-computed unimolecular rate constants, and fitted unimolecular rate constants for NH$_2$ addition to H$_2$CO on Mg(OH)$_2\cdot$H$_2$O.}
    \begin{tabular}{ccc}
        \hline
        \hline
        $T$ & $k$ & $k_{\mathrm{fit}}$ \\
        \hline
        10000.00  & 2.94 x 10$^{12}$ & {3.19 x 10$^{12}$} \\
        2265.62  & 1.11 x 10$^{12}$ & {1.18 x 10$^{12}$} \\
        1277.53  & 4.24 x 10$^{11}$ & {4.39 x 10$^{11}$} \\
        889.57  & 1.62 x 10$^{11}$ & {1.63 x 10$^{11}$} \\
        682.35  & 6.18 x 10$^{10}$ & {6.04 x 10$^{10}$} \\
        553.44  & 2.34 x 10$^{10}$ & {2.24 x 10$^{10}$} \\
        465.49  & 8.85 x 10$^{09}$ & {8.30 x 10$^{09}$} \\
        401.66  & 3.32 x 10$^{09}$ & {3.08 x 10$^{09}$} \\
        353.23  & 1.24 x 10$^{09}$ & {1.14 x 10$^{09}$} \\
        315.22  & 4.62 x 10$^{08}$ & {4.23 x 10$^{08}$} \\
        284.59  & 1.71 x 10$^{08}$ & {1.57 x 10$^{08}$} \\
        259.39  & 6.33 x 10$^{07}$ & {5.81 x 10$^{07}$} \\
        238.29  & 2.34 x 10$^{07}$ & {2.16 x 10$^{07}$} \\
        220.36  & 8.60 x 10$^{06}$ & {7.99 x 10$^{06}$} \\
        204.95  & 3.16 x 10$^{06}$ & {2.96 x 10$^{06}$} \\
        191.55  & 1.16 x 10$^{06}$ & {1.10 x 10$^{06}$} \\
        179.79  & 4.26 x 10$^{05}$ & {4.07 x 10$^{05}$} \\
        169.39  & 1.56 x 10$^{05}$ & {1.51 x 10$^{05}$} \\
        160.13  & 5.73 x 10$^{04}$ & {5.60 x 10$^{04}$} \\
        151.83  & 2.10 x 10$^{04}$ & {2.08 x 10$^{04}$} \\
        144.35  & 7.71 x 10$^{03}$ & {7.70 x 10$^{03}$} \\
        137.57  & 2.83 x 10$^{03}$ & {2.85 x 10$^{03}$} \\
        131.40  & 1.04 x 10$^{03}$ & {1.06 x 10$^{03}$} \\
        125.76  & 3.81 x 10$^{02}$ & {3.92 x 10$^{02}$} \\
        120.58  & 1.40 x 10$^{02}$ & {1.45 x 10$^{02}$} \\
        115.81  & 5.15 x 10$^{01}$ & {5.39 x 10$^{01}$} \\
        111.41  & 1.89 x 10$^{01}$ & {2.00 x 10$^{01}$} \\
        107.33  & 6.98 x 10$^{00}$ & {7.41 x 10$^{00}$} \\
        103.53  & 2.57 x 10$^{00}$ & {2.75 x 10$^{00}$} \\
        100.00  & 9.50 x 10$^{-01}$ & {1.02 x 10$^{00}$} \\
        \hline
        \hline
    \end{tabular}
    \tablefoot{The temperature $T$ is reported in K. The DFT-computed unimolecular rate constants $k$ and fitted unimolecular rate constants $k_{\mathrm{fit}}$ are reported in s$^{-1}$. The rate constants were computed at the M06-2X/pcseg-2 level of theory. The fitted values were obtained using the parameters reported in Table~\ref{tab:parameters} and Eq.~\ref{eq5}, with $\beta$ fixed to 0.}
    \label{tab:fitted_rates}
\end{table}

\begin{table}
    \centering
\caption{Temperature, thermal desorption rate constants, and desorption half-lives of NH$_2$ from the vdW complex on Mg(OH)$_2\cdot$H$_2$O.}
    \begin{tabular}{ccc}
        \hline
        \hline
       $T$ & $k_{\mathrm{des}}$ & $t_{1/2,\mathrm{des}}$ \\
        \hline
        10000.00  & {8.14 x 10$^{11}$} & {8.52 x 10$^{-04}$} \\
        2265.62  & {2.29 x 10$^{11}$} & {3.03 x 10$^{-03}$} \\
        1277.53  & {6.45 x 10$^{10}$} & {1.08 x 10$^{-02}$} \\
        889.57  & {1.81 x 10$^{10}$} & {3.82 x 10$^{-02}$} \\
        682.35  & {5.11 x 10$^{09}$} & {1.36 x 10$^{-01}$} \\
        553.44  & {1.44 x 10$^{09}$} & {4.82 x 10$^{-01}$} \\
        465.49  & {4.05 x 10$^{08}$} & {1.71 x 10$^{00}$} \\
        401.66  & {1.14 x 10$^{08}$} & {6.09 x 10$^{00}$} \\
        353.23  & {3.21 x 10$^{07}$} & {2.16 x 10$^{01}$} \\
        315.22  & {9.03 x 10$^{06}$} & {7.68 x 10$^{01}$} \\
        300.00  & {4.96 x 10$^{06}$} & {1.40 x 10$^{02}$} \\
        284.59  & {2.54 x 10$^{06}$} & {2.73 x 10$^{02}$} \\
        259.39  & {7.15 x 10$^{05}$} & {9.70 x 10$^{02}$} \\
        238.29  & {2.01 x 10$^{05}$} & {3.44 x 10$^{03}$} \\
        220.36  & {5.66 x 10$^{04}$} & {1.22 x 10$^{04}$} \\
        204.95  & {1.60 x 10$^{04}$} & {4.35 x 10$^{04}$} \\
        191.55  & {4.49 x 10$^{03}$} & {1.54 x 10$^{05}$} \\
        179.79  & {1.26 x 10$^{03}$} & {5.49 x 10$^{05}$} \\
        169.39  & {3.56 x 10$^{02}$} & {1.95 x 10$^{06}$} \\
        160.13  & {1.00 x 10$^{02}$} & {6.93 x 10$^{06}$} \\
        151.83  & {2.82 x 10$^{01}$} & {2.46 x 10$^{07}$} \\
        144.35  & {7.93 x 10$^{00}$} & {8.74 x 10$^{07}$} \\
        137.57  & {2.23 x 10$^{00}$} & {3.11 x 10$^{08}$} \\
        131.40  & {6.28 x 10$^{-01}$} & {1.10 x 10$^{09}$} \\
        125.76  & {1.77 x 10$^{-01}$} & {3.92 x 10$^{09}$} \\
        120.58  & {4.98 x 10$^{-02}$} & {1.39 x 10$^{10}$} \\
        115.81  & {1.40 x 10$^{-02}$} & {4.95 x 10$^{10}$} \\
        111.41  & {3.94 x 10$^{-03}$} & {1.76 x 10$^{11}$} \\
        107.33  & {1.11 x 10$^{-03}$} & {6.24 x 10$^{11}$} \\
        103.53  & {3.12 x 10$^{-04}$} & {2.22 x 10$^{12}$} \\
        100.00  & {8.80 x 10$^{-05}$} & {7.88 x 10$^{12}$} \\
        \hline
        \hline
    \end{tabular}
\tablefoot{The temperature $T$ is reported in K. The thermal desorption rate constants $k_{\mathrm{des}}$ are reported in s$^{-1}$, and the corresponding desorption half-lives $t_{1/2,\mathrm{des}}$ are reported in ns. The desorption rate constants refer to NH$_2$ desorption from the vdW complex and were evaluated using Eq.~\ref{eq:desorption}, with $E_{\mathrm{des}}=0.32$~eV computed at the M06-2X/pcseg-2 level of theory, corresponding to $E_{\mathrm{des},K}=3713.45$~K, and $\nu=1.18 \times 10^{12}$~s$^{-1}$.}
    \label{tab:NH2_desorption_rates}
\end{table}

\begin{table}
\centering
\caption{Temperature and forward-reaction half-lives for NH$_2$ addition to H$_2$CO on Mg(OH)$_2\cdot$H$_2$O.}
\begin{tabular}{cc}
\hline
\hline
$T$ & $t_{1/2}$ \\
\hline
10000.00 & 2.36 x 10$^{-13}$ \\
2265.62 & 6.23 x 10$^{-13}$ \\
1277.53 & 1.63 x 10$^{-12}$ \\
889.57 & 4.28 x 10$^{-12}$ \\
682.35 & 1.12 x 10$^{-11}$ \\
553.44 & 2.96 x 10$^{-11}$ \\
465.49 & 7.83 x 10$^{-11}$ \\
401.66 & 2.09 x 10$^{-10}$ \\
353.23 & 5.58 x 10$^{-10}$ \\
315.22 & 1.50 x 10$^{-09}$ \\
284.59 & 4.05 x 10$^{-09}$ \\
259.39 & 1.09 x 10$^{-08}$ \\
238.29 & 2.97 x 10$^{-08}$ \\
220.36 & 8.06 x 10$^{-08}$ \\
204.95 & 2.19 x 10$^{-07}$ \\
191.55 & 5.97 x 10$^{-07}$ \\
179.79 & 1.63 x 10$^{-06}$ \\
169.39 & 4.44 x 10$^{-06}$ \\
160.13 & 1.21 x 10$^{-05}$ \\
151.83 & 3.30 x 10$^{-05}$ \\
144.35 & 8.99 x 10$^{-05}$ \\
137.57 & 2.45 x 10$^{-04}$ \\
131.40 & 6.67 x 10$^{-04}$ \\
125.76 & 1.82 x 10$^{-03}$ \\
120.58 & 4.95 x 10$^{-03}$ \\
115.81 & 1.35 x 10$^{-02}$ \\
111.41 & 3.66 x 10$^{-02}$ \\
107.33 & 9.94 x 10$^{-02}$ \\
103.53 & 2.70 x 10$^{-01}$ \\
100.00 & 7.32 x 10$^{-01}$ \\
\hline
\hline
\end{tabular}
\tablefoot{The temperature $T$ is reported in K, and the corresponding forward-reaction half-life $t_{1/2}$ is reported in s. The half-life was calculated as $t_{1/2}=\ln 2/k$, where $k$ is the DFT-computed unimolecular rate constant reported in Table~\ref{tab:fitted_rates}.}
  \label{tab:half-life}
\end{table}
\clearpage
\section{DFT+U} \label{DFT+U}
{Different $U$ values were tested to identify the value that best reproduces the first barrier for NH$_2$ addition to H$_2$CO obtained at the CCSD(T)-F12/cc-pVDZ-F12//M06-2X/pcseg-1 level of theory, as reported in Table~\ref{tab:dft+U}.} The barriers for the $U$ value test were computed using the optimized geometry of the TS at the M06-2X/pcseg-1 level {and were referenced to the vdW complex, taken as zero. No BSSE correction was applied to these relative barrier heights because the vdW complex and the TS belong to the same local reaction segment.} 

{Since $U = 5$~eV reproduces the CCSD(T)-F12 barrier within 0.01~eV, higher $U$ values were not explored. Excessively large $U$ values may overcorrect electron localization and lead to unphysical results that deviate from the actual electronic structure. The $U = 5$~eV value was selected based on the first barrier and subsequently kept fixed for the entire reaction profile, without independent adjustment for the other stationary points. As discussed in Sect.\ \ref{benchmark}, the $U = 5$~eV correction improves the description of the overall reaction profile.}

\begin{table}[h]
    \centering
\caption{Hubbard $U$ parameters for NH$_2$ addition to H$_2$CO on Mg(OH)$_2\cdot$H$_2$O.}
    \begin{tabular}{@{}cc@{}}
        \hline\hline
   Method & 1st TS (eV) \\
        \hline
        CCSD(T)-F12/cc-pVDZ-F12  & 0.27 \\
       M06-2X/pcseg-2 & 0.23 \\
        PBE+U1/TZP  & 0.10 \\
        PBE+U2/TZP  & 0.15 \\
        PBE+U3/TZP  & 0.19 \\
        PBE+U4/TZP  & 0.23 \\
        PBE+U5/TZP  & 0.26 \\
        \hline\hline
    \end{tabular}
\tablefoot{The Hubbard $U$ parameters were tested using PBE+$U$/TZP single-point calculations on M06-2X/pcseg-1 optimized geometries for the first TS (see Fig.~\ref{fgr:PES}). CCSD(T)-F12 and M06-2X reference values are also reported. The barrier heights are reported in eV and referenced to the vdW complex, taken as zero.}
    \label{tab:dft+U}
\end{table}

\end{appendix}

\end{document}